\documentclass[sigconf]{acmart}

\AtBeginDocument{%
  \providecommand\BibTeX{{%
    \normalfont B\kern-0.5em{\scshape i\kern-0.25em b}\kern-0.8em\TeX}}}

\setcopyright{acmcopyright}
\copyrightyear{2025}
\acmYear{2025}

\acmConference[C\&C '25]{C\&C '25 ACM Creativity and Cognition 2025 conference}{June 23--25, 2025}{Virtual}
\acmBooktitle{C\&C '25: ACM Creativity and Cognition 2025 conference,
  June 23--25, 2025, Virtual}
\acmPrice{15.00}
\acmISBN{978-1-4503-XXXX-X/18/06}

\newcommand{\re}[1]{{\color{black} #1}}

\usepackage{multirow}
\usepackage{subcaption}
\usepackage{caption}

\begin{document}



\title[Structuring GenAI-assisted Hypotheses Exploration with an Interactive Shared Representation]{``The Diagram is like Guardrails": \\Structuring GenAI-assisted Hypotheses Exploration with an Interactive Shared Representation}


\author{Zijian Ding}
\authornote{This work was conducted during the author's internship at IBM Research.}
\affiliation{%
  \institution{College of Information, University of Maryland}
  \country{USA}}
  
\author{Michelle Brachman}
\affiliation{%
  \institution{IBM Research}
  \country{USA}}
  
\author{Joel Chan}
\affiliation{%
  \institution{College of Information, University of Maryland}
  \country{USA}}
  
\author{Werner Geyer}
\affiliation{%
  \institution{IBM Research}
  \country{USA}}

\renewcommand{\shortauthors}{Ding, et al.}

\begin{abstract}
Data analysis encompasses a spectrum of tasks, from high-level conceptual reasoning to lower-level execution. While AI-powered tools increasingly support execution tasks, there remains a need for intelligent assistance in conceptual tasks. This paper investigates the design of interactive tree diagrams as effective shared representations for AI-assisted hypothesis exploration. We developed a system with ordered node-link diagram augmented with AI-generated information hints and visualizations. Through a design probe (n=22), participants generated diagrams averaging 21.82 hypotheses. Our findings showed that the node-link diagram acts as ``guardrails" for hypothesis exploration, facilitating structured workflows, providing overviews, and enabling backtracking. The AI-generated information hints, particularly visualizations, aided users in transforming abstract ideas into data-backed concepts while reducing cognitive load. We further discuss how node-link diagrams can support both parallel exploration and iterative refinement in hypothesis formulation, potentially enhancing the breadth and depth of human-AI collaborative data analysis.
\end{abstract}

 




\begin{teaserfigure}
  \includegraphics[width=\textwidth]{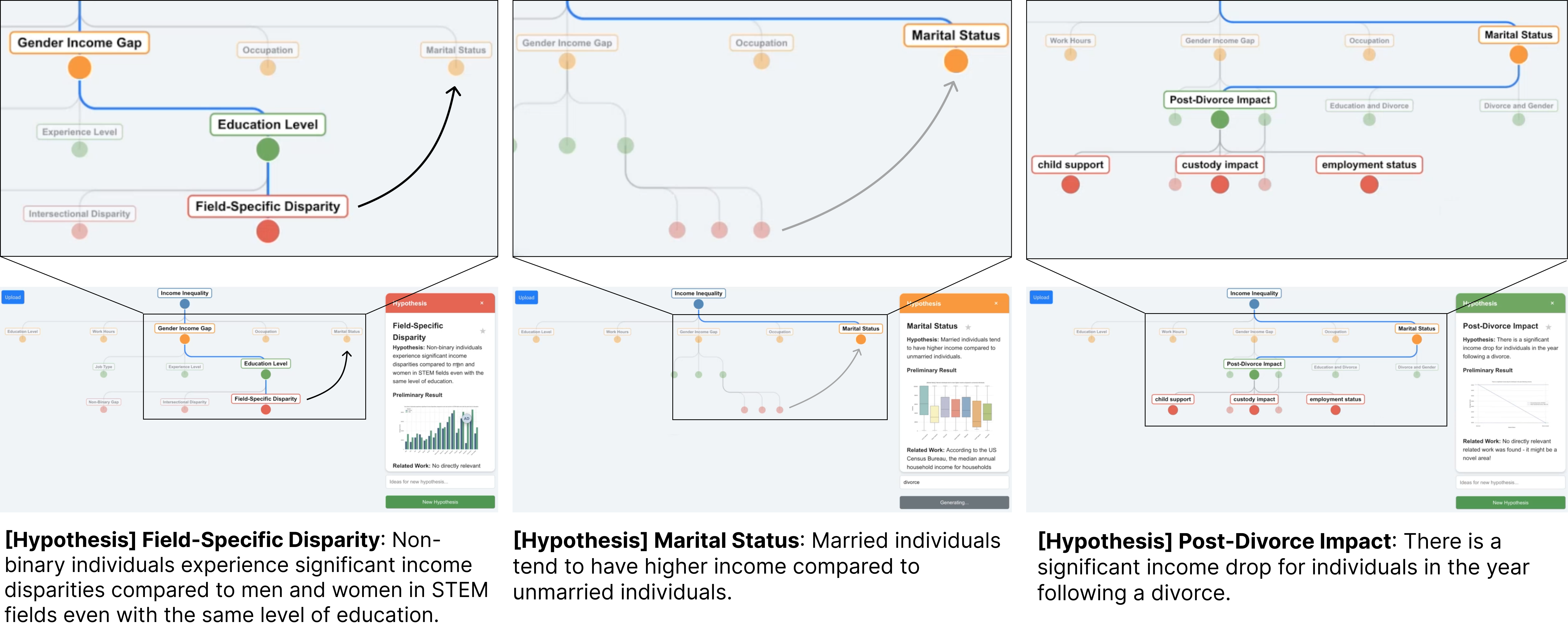}
  \caption{Our system supports nonlinear AI-assisted hypothesis exploration that balances breadth and depth of exploration. Using the node-link diagram shared representation and integrated information hint panels that display preliminary results and related work, a participant in our user study deeply explored a branch of hypotheses around gender income gaps, including a nuanced hypothesis about field-specific gender disparities in income (A); but also simultaneously kept track of the overall hypothesis space, and backtracked to explore other hypothesis branches around variations in income by marital status (B), such as post-divorce impacts on income. A more detailed version Figure \ref{fig:P3} can be found in Appendix \ref{appendix:B}.}
  \Description{Ordered node-link diagram for data analysis hypothesis exploration by supporting backtracking for exploration and iteration.}
  \label{fig:teaser}
\end{teaserfigure}

\maketitle

\section{Introduction}

Data analysis is a complex task that involves many subtasks ranging from high-level conceptual reasoning and planning to lower-level execution tasks. High-level tasks include checking domain assumptions, thinking through confounds, and identifying interesting hypotheses to test \cite{junHypothesisFormalizationEmpirical2022}. Lower-level tasks, which can sometimes be tedious, include implementing statistical models or producing specific data visualizations to test domain assumptions \re{\cite{kazemitabaarImprovingSteeringVerification2024}}.

As Generative AI (GenAI) models continue to advance, there are an increasing number of GenAI-powered tools supporting the \textit{execution} of specific data analysis tasks, such as creating visualizations. For instance, tools like LIDA \cite{dibiaLIDAToolAutomatic2023} and ChartGPT \cite{tianChartGPTLeveragingLLMs2024} streamlined the process of generating data visualizations. \re{Researchers also developed interactive task decomposition \re{tools} to steer and verify executions in LLM-Assisted data analysis \cite{kazemitabaarImprovingSteeringVerification2024}.} Meanwhile, there is a growing interest in providing intelligent support for the higher-level conceptual tasks of data analysis. For example, rTisane \cite{junRTisaneExternalizingConceptual2024} was designed to support domain assumption checking in data analysis with GenAI. Even at the model level, Large Concept Models \cite{teamLargeConceptModels2024} have emerged, offering new approaches to higher-level conceptual understanding and generation by operating at a sentence-level semantic representation space.

A direct approach to supporting these higher-level tasks might be to deploy GenAI tools in an autonomous, end-to-end manner, to both produce a large number of possible hypotheses and select appropriate ones to pursue \cite{zhouHypothesisGenerationLarge2024,tong2024automating,siCanLLMsGenerate2024}. \re{However, GenAI-driven evaluation and selection of hypotheses remains an open problem such as novelty assessment of hypotheses \cite{siCanLLMsGenerate2024}.} A core barrier here is that hypothesis evaluation and selection require judgment with expert domain knowledge of what is more or less interesting or relevant. For instance, in a study on income inequality, a GenAI tool might generate hypotheses about correlations between various demographic factors and income. However, it may not have the nuanced understanding to prioritize hypotheses that are most relevant to current socioeconomic debates or policy considerations. Although such knowledge is pervasive in human society, it often exists in tacit forms, undocumented and unpublished \cite{ackermanSharingKnowledgeExpertise2013,cetinaEpistemicCulturesHow1999,rittbergEpistemicInjusticeMathematics2018}, making it difficult to transfer this decision-making ability to artificial systems.

Moreover, the process of hypothesis exploration requires iterative steering towards deeper, more nuanced insights. This involves increasing both the breadth (exploring various facets of the problem) and depth (diving into specific areas of interest) of exploration. For example, the field of Alzheimer's research recently faced a controversy surrounding the long-held amyloid hypothesis \cite{AlzheimersResearchUK2022}. This hypothesis, which posits that a protein called amyloid is a major contributor to brain dysfunction in Alzheimer's disease, has influenced research for decades. Recent controversies have prompted researchers to broaden their scope to include factors like tau proteins and inflammation, while deepening their understanding of these elements' complex interactions \cite{AlzheimersResearchUK2022}. This shift highlights the importance of exploring alternative hypotheses (breadth) and critically examining long-standing assumptions (depth) through a non-linear exploration process. Given these challenges, we argue that effective hypothesis exploration in data analysis should be a \textit{mixed-initiative} \cite{horvitzPrinciplesMixedinitiativeUser1999} process that integrates the complementary strengths of human and AI capabilities. This leads us to the question: How can we support human-AI collaborative exploration of hypotheses?

A common design pattern in mixed-initiative systems is a shared representation \cite{heerAgencyAutomationDesigning2019}, which allows AI systems to computationally analyze and understand human tasks, while also providing interfaces through which humans can interact with, evaluate, modify, or reject AI-generated suggestions. For instance, DirectGPT \cite{leeDesignSpaceIntelligent2024} uses a shared representation of text, code, and images that both the AI and human user can manipulate through a direct manipulation interface, transforming user actions into engineered prompts. Similarly, in collaborative writing systems, shared document representations allow both AI writing assistants and human authors to contribute to and refine the text \cite{leeDesignSpaceIntelligent2024}. 
In AI-assisted data analysis, shared representations have been developed to support subtasks such as data cleaning \cite{kandelWranglerInteractiveVisual2011,guoProactiveWranglingMixedinitiative2011} and data visualization \cite{wongsuphasawatVoyagerExploratoryAnalysis2016,wongsuphasawatVoyagerAugmentingVisual2017,satyanarayanVegaLiteGrammarInteractive}. But support for the messier, conceptual work of hypothesis exploration --- in a way that helps ensure sufficient breadth and depth of exploration --- remains underexplored. Thus, in this paper, we ask the following research question: \textbf{How might we design effective shared representations for GenAI-assisted hypothesis exploration?} 

Our research addresses this question through the following contributions:

\begin{itemize}
    \item \textbf{System design}: We designed and developed a GenAI-assisted ordered node-link diagram as a potential structured shared representation for hypothesis exploration. This tool interprets users' analytical intents, generates potential hypotheses and corresponding visual hypotheses, and retrieves relevant literature as supporting texts.
    \item \textbf{Empirical results}: Using this prototype tool as a design probe, our qualitative study with 22 data analysts suggests that the analysts find the node-link diagram helpful as a shared representation to structure exploration of and iteration on AI-generated hypothses, and the information hints, especially visual hypotheses, can help guide analysis decisions.
\end{itemize}

\section{Related Work}

To inform our system design that supports both breadth-oriented exploration and depth-oriented iteration in hypothesis formulation, we reviewed existing GenAI-assisted tools for data analysis and coding, highlighting their focus on task execution rather than conceptual exploration. We then examined research on shared representations between humans and AI, particularly the use of node-link diagrams for providing overviews and supporting sense-making. We also explored studies on multimodal information hints that combine textual and visual data to inform data analysis.

\subsection{GenAI-Assisted Data Analysis Interfaces: From Task Execution to Conceptual Exploration}

Data analysis, as a discipline with mature methodologies, has a vast amount of relevant information available online that has been absorbed by Large Language Models (LLMs), enabling automated execution of specific data analysis subtasks, such as generating data visualizations \cite{dibiaLIDAToolAutomatic2023}. LIDA, functioning as a backend engine, can automatically analyze datasets, generate data analysis goals, and produce visualization code and visualizations \cite{dibiaLIDAToolAutomatic2023}. Leveraging the data analysis capabilities of LLMs, multiple GenAI-assisted data analysis interfaces have been developed \cite{dingIntelligentCanvasEnabling2024,vaithilingamDynaVisDynamicallySynthesized2024,weideleEmpiricalEvidenceConversational2024}. While these automated approaches demonstrate the potential of GenAI in data analysis, there is also a need for mixed-initiative \cite{horvitzPrinciplesMixedinitiativeUser1999} interaction design to allow users to actively steer and iterate on results produced by GenAI, as opposed to passively accepting the results of an end-to-end process. DynaVis \cite{vaithilingamDynaVisDynamicallySynthesized2024} generates data visualizations through a single command and provides an adaptive interface for refining the visualization. Some approaches combine chatbots with graphical user interfaces (GUIs), synchronizing chatbot interactions with corresponding GUI supports for linear, one-way investigation of data \cite{weideleEmpiricalEvidenceConversational2024}. 
However, these GenAI interfaces primarily focus on executing analysis procedures with technical precision - ``doing things right" - rather than making context-appropriate decision as ``doing the right thing". \re{Some prior work suggests desire paths for GenAI-assisted approaches that extend beyond execution, to assist with more conceptual tasks like planning data analysis  \cite{guHowDataAnalysts2024}. Other conceptual tasks may include exploratory data analysis (EDA) to probe and understand structures and patterns in the data to inform planning and hypothesizing. In this work, we build specifically on Jun et al's modeling of hypothesis formalization as a key subtask of data analysis: iterating over the space of conceptual hypotheses in conversation with properties of the data to formalize hypotheses into statistical models and procedures that can then be executed \cite{junHypothesisFormalizationEmpirical2022,junRTisaneExternalizingConceptual2024}. This subtask of hypothesis formalization differs from EDA in that it explicitly focuses on exploring and formalizing \textit{conceptual hypotheses} (which can often be stated in natural language) rather than broadly probing for structures and patterns with visualizations of data distributions and properties.}


\subsection{Shared Representation for Data Analysis and Coding}

A shared representation between humans and AI constitutes an interface that both can manipulate concurrently in a task-appropriate manner. As AI capabilities progressively advance, systems become increasingly capable of accurately discerning and even anticipating user intentions within the shared representation \cite{heerAgencyAutomationDesigning2019}. An example is data wrangling: when a user selects a word in the initial row, the system responds by proposing automatic transformations, such as extracting or truncating text from table cells \cite{kandelWranglerInteractiveVisual2011,guoProactiveWranglingMixedinitiative2011}. In those work, the table serves as a shared representation, allowing the AI to suggest actions and humans to evaluate and refine them. This approach enhances interaction efficiency and accuracy. Alternative methods like textual exchanges or command sequences would make evaluation and iteration more difficult, often obscuring task structure or action outcomes. In AI-assisted code generation, the AI first sketches a code outline based on the user's prompt before generating the complete code \cite{zhu-tianSketchThenGenerate2024}. Similarly, in frontend code generation, users can sketch the website layout, prompting the AI to generate a corresponding product requirement document, which subsequently informs the code generation process \cite{zhangFrontendDiffusionExploring2024}. For data analysis, shared representations have evolved from traditional notebooks and dashboards to more sophisticated forms such as canvases \cite{dingIntentbasedUserInterfaces2024}, as well as language and block-based interfaces \cite{caoDataParticlesBlockbasedLanguageoriented2023}.
An important consideration for shared representations in data analysis is the provision of an overview that contextualizes each instance, facilitating sense-making \cite{chanFormulatingFixatingEffects2024}. Node-link diagrams, which have been increasingly adopted as a shared representation for Generative AI \cite{jiangGraphologueExploringLarge2023,huangCausalMapperChallengingDesigners2023}, show promise as an overview mechanism, elucidating relationships between instances \cite{chanFormulatingFixatingEffects2024}. Currently, node-link diagrams serve as shared representations for data analytics workflows in systems such as WaitGPT \cite{xieWaitGPTMonitoringSteering2024} and commercialized products such as Altair RapidMiner\footnote{https://altair.com/altair-rapidminer} and KNIME\footnote{https://www.knime.com/}. 
Those applications enable visual coordination of specific subtasks of executing a data analysis workflow, such as moving through data cleaning and executing specific sequences of models. \re{We build on these prior explorations to investigate the potential of a node-link diagram pattern as a shared representation for supporting GenAI-assisted hypothesis exploration. We hypothesize that the structure of node-link diagrams could facilitate their use as shared representations given their alignment with the multi-thread representation paradigm of multiverse analysis \cite{dragicevicIncreasingTransparencyResearch2019a}.}

\subsection{Multimodal Information Hints for Data Analysis}

Effective hypothesis exploration often extends beyond text-based approaches, requiring data scientists and researchers to balance reasoning about conceptual hypotheses with consideration of the underlying data, and statistical models that connect these hypotheses with the data; thus, contextual signals from the underlying data and models are needed to gain sufficient context to decide on analysis directions \cite{junHypothesisFormalizationEmpirical2022}. In this paper, we refer to these valuable contextual signals as ``information hints" for brevity, drawing on insights from Information Foraging Theory \cite{pirolliInformationForaging1999}, which conceptualizes users as foragers following ``information scents" to navigate complex information landscapes and balance search costs against potential benefits.  For text-based information hints, Chen et al.'s ``sketch-then-generate" technique \cite{zhu-tianSketchThenGenerate2024} creates incomplete code plans during the prompt crafting phase, followed by the generation of complete code. In Retrieval-augmented generation (RAG) \cite{lewisRetrievalAugmentedGenerationKnowledgeIntensive2020} query-relevant data sources are retrieved to contextualize the question. In the visual domain, GenAI tools like LIDA \cite{dibiaLIDAToolAutomatic2023} automate the generation of data visualizations, enabling data analysts to engage with and derive patterns from datasets more efficiently. Bao et al. \cite{baoRecommendationsVisualizationRecommendations2022} and Quadri et al. \cite{quadriYouSeeWhat2024} have focused on enhancing comprehension through visual recommendations and elicitation techniques. Bridging textual and visual modalities, Zheng et al. \cite{zheng2022telling} explored blending text and visuals in data analysis slides, while Chen et al. \cite{chenCrossDataLeveragingTextData2022} developed methods to connect text with visualizations and tables. We build on those previous works on information hints to support the decision-making process in hypothesis exploration such as judging whether a hypothesis candidate is worth pursuing or not.

\section{System Design}

As discussed previous sections, hypothesis exploration needs balancing breadth and depth while allowing for nonlinear exploration paths as a cornerstone of conceptual analysis. However, existing research lacks a clear framework for designing interfaces that concretely facilitate such nonlinear exploration. Therefore, we conducted a formative study to gather insights and explore the specific needs and expectations of data scientists in designing interfaces that support hypothesis exploration.

\subsection{Formative Study}

\begin{figure*}
\includegraphics[width=0.9\textwidth]{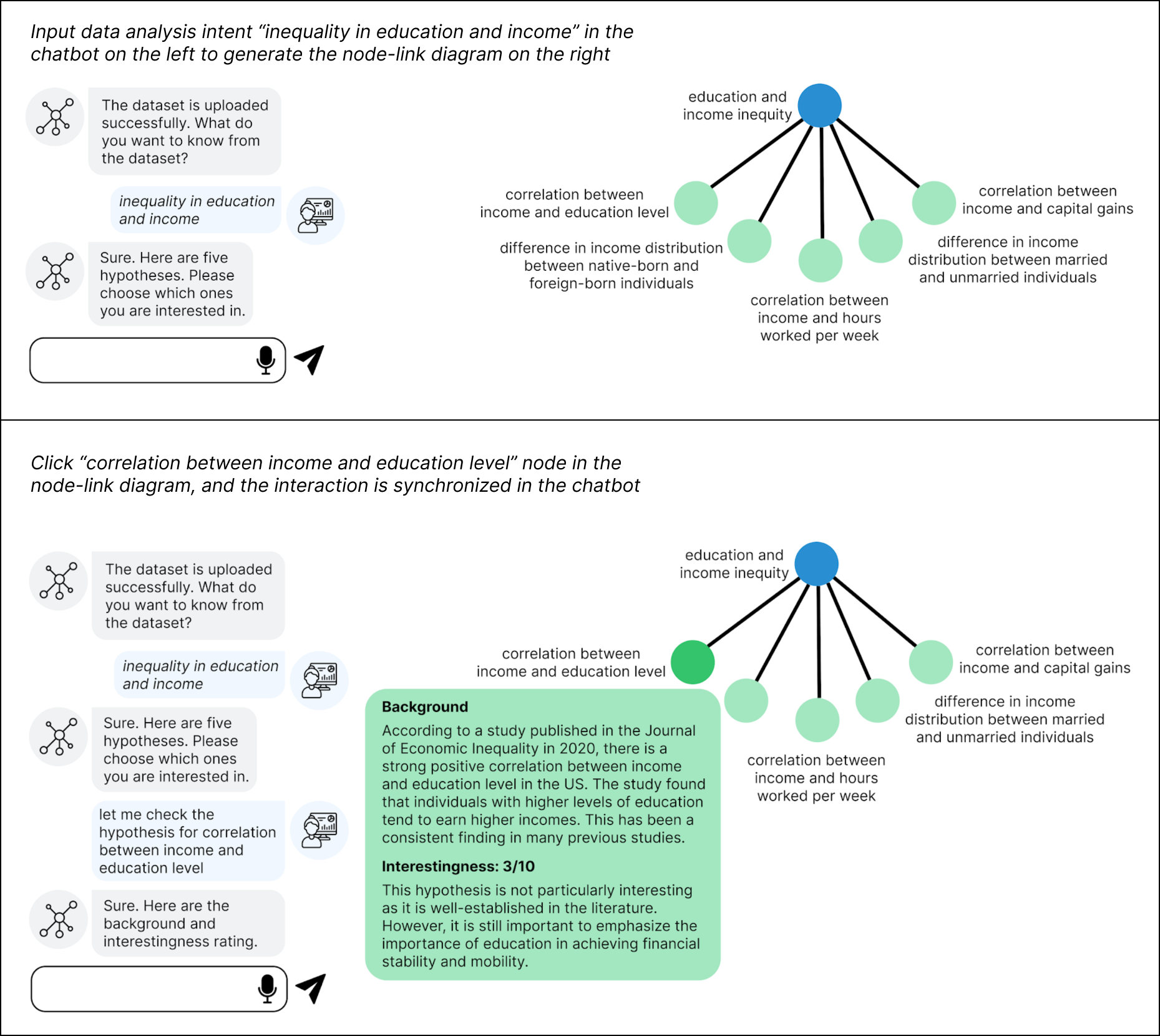}
  \caption{The wireframes depict the initial design for the formative study, showcasing synchronizing a chatbot with a node-link diagram based on \cite{weideleEmpiricalEvidenceConversational2024}. In the upper figure, after the user uploads a dataset and inputs the analysis intent ``inequality in education and income" in the chatbot, the system auto-generates five hypotheses (e.g., ``correlation between income and education level") in the node-link diagram. The lower figure demonstrates interactivity: when a user clicks a hypothesis node, such as ``correlation between income and education level," the diagram shows the hypothesis background and an interestingness rating, while simultaneously updating the chatbot with a message ``let me check the hypothesis for correlation between income and education level."}
  \Description{The wireframes depict the initial design for the formative study, showcasing synchronizing a chatbot with a node-link diagram based on \cite{weideleEmpiricalEvidenceConversational2024}. In the upper figure, after the user uploads a dataset and inputs the analysis intent ``inequality in education and income" in the chatbot, the system auto-generates five hypotheses (e.g., ``correlation between income and education level") in the node-link diagram. The lower figure demonstrates interactivity: when a user clicks a hypothesis node, such as ``correlation between income and education level," the diagram shows the hypothesis background and an interestingness rating, while simultaneously updating the chatbot with a message ``let me check the hypothesis for correlation between income and education level."}
  \label{fig:initialdesign}
\end{figure*}

To inform our system design, we conducted a formative study with six participants (2 research scientists and 4 data science interns), gathering feedback on initial wireframe prototypes (see Figure \ref{fig:initialdesign} as an example). Our formative study centered on three primary aspects: 
(1) evaluating different shared representations to support structured exploration and iteration, and asking participants' desired functions and interactions such as ``Will you navigate back the network?" (2) comparing single-thread chatbot interfaces versus multi-thread node-link diagram interfaces, and (3) identifying the appropriate combination of information hints and contextual details to display for each hypothesis to support user analysis needs.

The formative study revealed a preference for a non-chatbot interface as a shared representation. This preference stemmed from concerns that a chatbot-style interaction might inadvertently constrain the exploratory nature of hypothesis exploration. For instance, formative study participant F1 noted that the hypotheses exploration was still linear because the chatbot view limits the exploration. This insight suggested that supporting non-linear exploration of hypotheses would be more beneficial in the initial stages, rather than focusing on synchronizing conversation with a chatbot interface.

Regarding information hints to evaluate the potential of hypotheses, our findings indicated a strong preference among data scientists for early access to statistical data. F1, F2 and F3 emphasized the importance of engaging directly with the dataset from the outset of the exploration process. Subsequent feedback advocated for the inclusion of a statistical overview or preliminary data analysis and the background of the hypothesis to provide an overview of each hypothesis: ``One of the first thing is that you do some of overview, analysis of your dataset...I don't just wanna know what the
possible hypothesis are, right? I wanna know, also, good overview
of the data." (F1)

However, this approach introduced new challenges, particularly in terms of information density and display. \re{While F3 hovered over the node at the bottom, they observed that there was too much information to fit on the screen as the exploration got deeper}, highlighting the limitations of traditional detail-on-demand interaction in this context. To address this issue, participants proposed alternative solutions such as implementing a table and summary view, or potentially opening a new page for more detailed information.


\subsection{Design Goals and Implementation}

In summary, our reflections on prior research on hypothesis exploration, and findings from our formative study, led us to formulate the following three design goals:

\begin{enumerate}
    \item \textbf{DG1: Enable users to balance breadth and depth of exploration.} The system should offer a clear, scalable structure that showcases the breadth of potential hypotheses while facilitating detailed investigation of each. This design goal was primarily motivated by prior research \cite{junHypothesisFormalizationEmpirical2022}.
    \label{dg:1}
    \item \textbf{DG2: Support non-linear exploration of hypotheses}. The system should support revisiting, backtracking, or branching off from existing hypotheses to accommodate the iterative nature of hypothesis exploration. This design goal emerged in part from prior research \cite{guUnderstandingSupportingDebugging2023}, and was refined by the findings from the formative study to remove the chatbot interface which limited the multi-thread exploration.
    \label{dg:2}
    \item \textbf{DG3: Enable integrated engagement with relevant data while exploring hypotheses.} The system needs to integrate data views and textual insights to maintain context-driven analysis within the hypothesis space. This design goal emerged primarily from the formative study insights.
    \label{dg:3}
\end{enumerate}



\begin{figure*}
\includegraphics[width=1\textwidth]{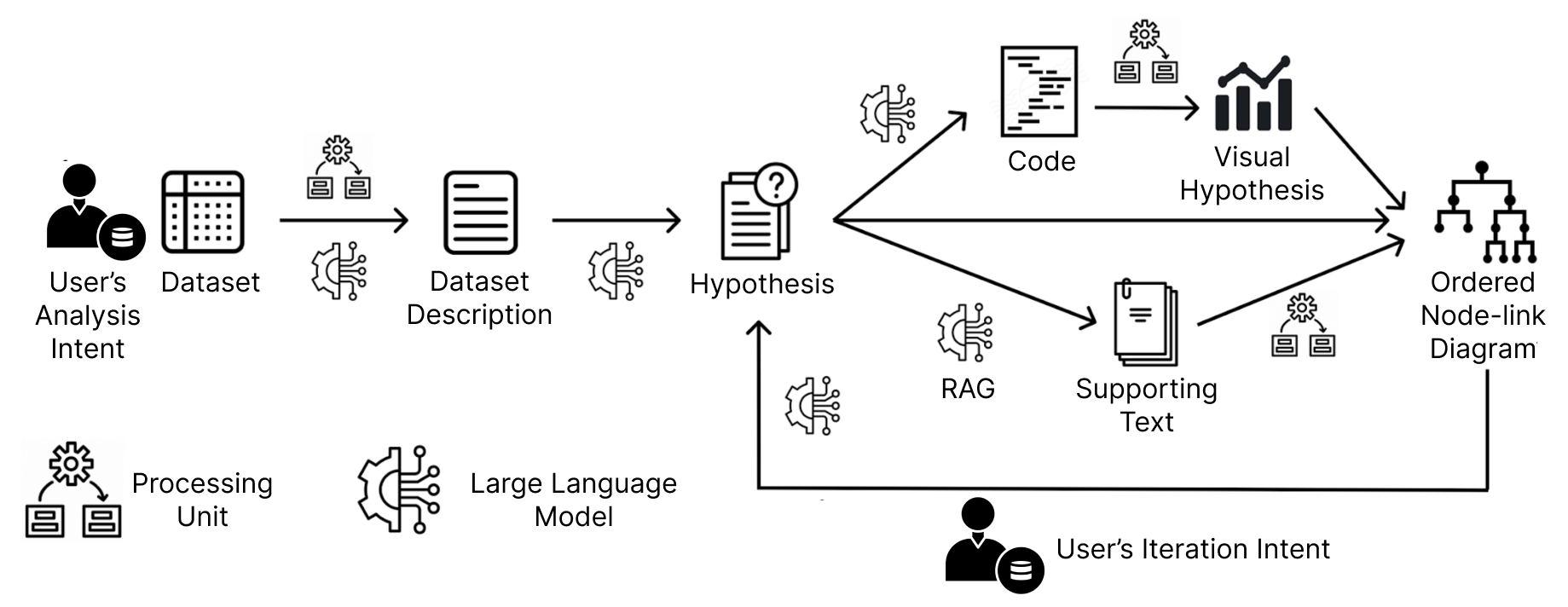}
  \caption{System architecture illustration of the AI-powered hypothesis exploration system.}
  \Description{}
  \label{fig:system-architecture}
\end{figure*}

To address our design goals, we designed and developed a hypothesis exploration system to assist users in exploring hypotheses more effectively. The system architecture, illustrated in Figure \ref{fig:system-architecture}, comprises three components: Intent-Hypothesis Interpreter, Ordered Node-link Diagram Visualizer, and Information Hints Explorer. The language model utilized for this system was GPT-4o\footnote{https://platform.openai.com/docs/models/gpt-4o}, accessed in July 2024.

\subsubsection{Intent-Hypothesis Interpreter}

The user initiates the hypothesis exploration process by inputting a high-level analysis intent, such as ``analyze the dataset from the perspective of income inequality." In response, the system generates and visually displays an initial set of potential hypotheses for the user to explore. The interpreter also accepts iteration intents and generates new hypothesis branches. The hypothesis generation is accomplished through a large language model that analyzes the data, employing a chain of prompts to extract information about the input data (requiring semantically meaningful column names and metadata), and then using this information along with user inputs to generate potential hypotheses based on its training data and provided context. The intent-hypothesis interpreter is created based on the data summary function of LIDA \cite{dibiaLIDAToolAutomatic2023}, by adding a focus on hypothesis exploration. The interpreter also accepts iteration intents (i.e., intents to elaborate or branch off from an existing hypothesis), and generates new hypothesis candidates that branch off from a specific hypothesis (\textbf{DG\ref{dg:1}}). The prompts for initial hypothesis generation and new hypothesis branch generation (hypothesis iteration) can be found in Appendix \ref{appendix:A}.
    
\subsubsection{Ordered Node-link Diagram Visualizer}

The LLM-generated multi-thread hypotheses were presented in a node-link diagram as a shared representation for exploring and iterating on hypotheses. To support balancing breadth and depth of exploration (\textbf{DG\ref{dg:1}}), we employed a stratified approach to hypothesis generation: initially presenting five hypotheses, followed by three hypotheses for subsequent iterations. Our reasoning was that three to five alternatives might cover the main possibilities in each iteration, and too many long-tail hypotheses would yield diminishing marginal returns.

The tree-like semantics of the diagram constitutes a shared representation of breadth and depth of exploration. We adopted a version control view approach \cite{manesh2024sharp}, using vertical positioning to represent hypothesis iterations and horizontal branching to show alternative exploration paths. Thus, the horizontal width and number of branches (along with their labels) helps the analyst keep a holistic grasp of the ongoing breadth of exploration and diversity of hypothesis under consideration. At the same time, the analyst can perceive and control depth of exploration towards successively more complex or elaborated hypotheses by drilling down a specific node to generate more branches, while retaining a sense of the hierarchy of previous parent hypotheses for a given hypothesis node. 
To better balance breadth and depth, we reasoned that it is important for the analyst to be able to see many hypothesis branches at once. Thus, to ensure the diagram fits within the screen, we implemented a position adjustment mechanism. This algorithm identified the leftmost and rightmost nodes at each level, then adjusted child node positions based on their parent's location. For edge nodes, children were aligned to the corresponding side, while for central nodes, children were repositioned towards their average position. This approach helped prevent overlap and created a more balanced layout, optimizing screen space utilization without losing the logical structure of the hypotheses. In this way, the semantics and affordances of the node-link diagram can support the user in balancing breadth and depth of exploration (\textbf{DG\ref{dg:1}}). 

Enabling visibility of the macro structure of the hypothesis tree also enables analysts to explore the hypothesis space not just by considering the most adjacent hypotheses, but also with more distant moves in the hypothesis space that are more characteristic of the nonlinear nature of hypothesis exploration, such as backtracking to a hypothesis further up in the hierarchy from a different branch (\textbf{DG\ref{dg:2}}).
    
\subsubsection{Information Hints Explorer}

Each hypothesis was accompanied by information hints. The hints included a visual hypothesis as a preliminary view of the data involved and supporting text around the hypotheses from Wikipedia sources, if available. We generated the visual hypothesis with LIDA \cite{dibiaLIDAToolAutomatic2023} and retrieved supporting text with a Retrieval-Augmented Generation (RAG) API service. This information is displayed in the sidebar of the interface when a given hypothesis node is clicked on, enabling integrated engagement with data while exploring hypotheses (\textbf{DG\ref{dg:3}}).

\subsection{Usage Scenario}

\begin{figure*}
\includegraphics[width=1\textwidth]{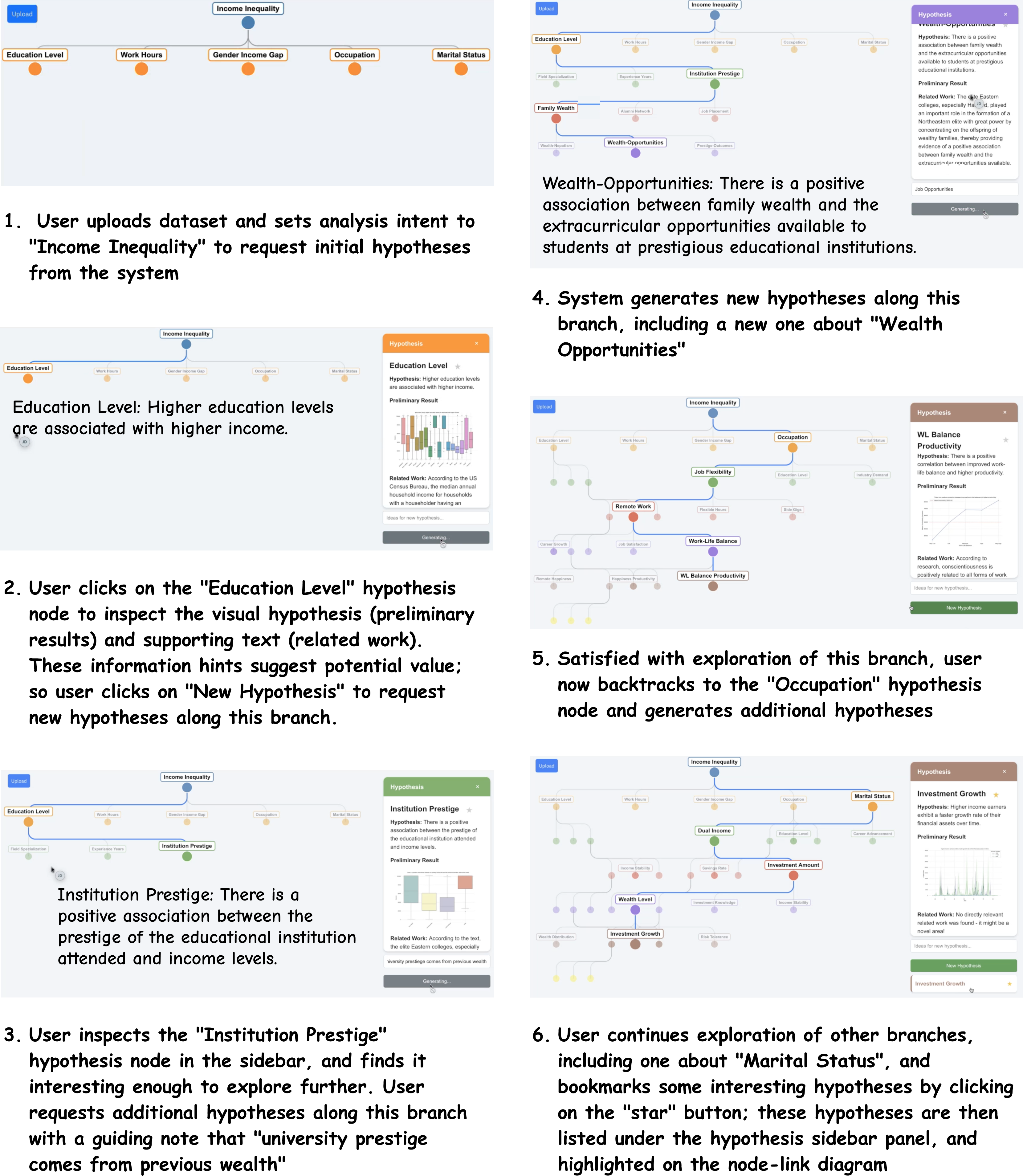}
  \caption{Walk-through of hypothesis exploration functions and P5's behaviors.}
  \Description{}
  \label{fig:user-journey}
\end{figure*}


To illustrate how participants actually engaged with the hypothesis exploration tool, Figure \ref{fig:user-journey} offers a walk-through of functionality (based on P5's exploration process). P5 started with uploading a 2020 US census dataset and entered the analysis intent ``income inequality" to generate initial hypotheses. P5 then interacted with the system by clicking on hypothesis nodes to examine information hints, including visual hypotheses and supporting text. P5 began by examining Education Level and narrowed down to Institution Prestige hypothesis \textit{There is a positive association between the prestige of the educational institution attended and income levels.} Next, P5 manually added the idea ``University prestige comes from previous wealth," which led the system to generate the Family Wealth hypothesis \textit{There is a positive association between family wealth and the prestige of the educational institution attended.} Building on this, P5 introduced the thought ``Nepotism is consistent within schools with more prestige and families that have increased wealth," prompting the generation of a Wealth-Opportunities hypothesis \textit{There is a positive association between family wealth and the extracurricular opportunities available to students at prestigious educational institutions.} After having sufficiently explored this hypothesis branch, the user can generate additional branches by backtracking to ``Occupation" node. During the process, participants could expand visual hypotheses for deeper analysis and bookmarked a hypothesis, which was then emphasized in the diagram and listed in the hypothesis panel.

\section{Study Design}
\label{sec:study_design}

Given the system's focus on supporting new forms of AI-assisted hypothesis exploration rather than optimizing speed or accuracy on a predefined analytics task, we focused on gathering qualitative insights into how participants actually used the interface, whether and how they perceived its potential value, and what pitfalls emerged in usage scenarios. Our goal at this stage was not to measure ``efficacy" in a strict experimental sense, such as comparing time-to-insight or user accuracy against a baseline tool, but rather to understand the user experience in depth and uncover design implications.

We thus designed our study to answer the following three research questions (each targeting one of our core design goals):

\begin{enumerate}
    \item \textbf{How did users balance breadth and depth of exploration?} (DG\ref{dg:1}) We wanted to know whether participants could have an overview of hypotheses and then selectively drill down to examine any one in depth. Therefore, we observed how many hypotheses were generated, how participants navigated the node-link diagram, and in what ways they perceived the approach as supportive of both global overviews (breadth) and focused investigations (depth).

    \item \textbf{How did users engage in non-linear exploration of hypotheses?} (DG\ref{dg:2}) We were interested in how participants' used the system to engage in non-linear exploration behavior such as backtracking other branches. We also asked participants to compare this non-linear process with the linear iteration process, such as using chatbots.

    \item \textbf{How did users engage with relevant data while exploring hypotheses?} (DG\ref{dg:3}) We also asked participants how they used the ``information hints" such as visual hypotheses, and whether those hints facilitate exploring hypotheses.
\end{enumerate}

Because these research questions were mainly qualitative---focusing on participant experiences rather than strict performance metrics---interviews allowed us to probe the why and how behind participant behaviors. In the interview, participants were asked to reflect on the study processes, as well as the perceived benefits and challenges, which are detailed in Section \ref{sec:study_procedure}. We further triangulated this interview data with our detailed interaction logs and the final node-link diagrams they generated. 

In summary, we collected and analyzed the following data:

\begin{itemize}
    \item Resulting Diagram Artifacts: Each participant's final diagram structure (breadth, depth, number of branches and nodes) offered a measurement of how their exploration evolved.
    \item Interaction Logs: Detailed records of how participants navigated the node-link diagram, which nodes they expanded or branched from, and the sequence of their exploration.
    \item Qualitative Interviews: In-depth conversations about participants' experiences, including perceived utility, ease of navigation, pitfalls.
\end{itemize}


\subsection{Participants}

To ensure a wide range of perspectives, we recruited 22 participants within a multinational technology corporation with diverse professional backgrounds in data analysis. The participant pool consisted of 7 data scientists or data-informed consultants, 8 software developers, 3 researchers, and 4 business professionals. Table \ref{tab:participant-demographics} provides a detailed breakdown of participant demographics.

\begin{table}[htbp]
\caption{Participants' demographic information including occupation and location.}
    \label{tab:participant-demographics}
\begin{tabular}{|c|c|c|}
\hline
ID & Occupation & Location \\
\hline
P1 & Research Scientist & North America \\
P2 & Developer & North America \\
P3 & Global Sales & North America \\
P4 & Business Technology Leadership & North America \\
P5 & Developer & North America \\
P6 & Consultant & Asia-Pacific \\
P7 & Developer & Asia-Pacific \\
P8 & Data Scientist & Asia-Pacific \\
P9 & Developer & Asia-Pacific \\
P10 & Developer & North America \\
P11 & Developer & North America \\
P12 & UX Researcher & Europe \\
P13 & Consultant & North America \\
P14 & Research Engineer & North America \\
P15 & Developer & North America \\
P16 & Learning Content Development & North America \\
P17 & Finance and Operations & North America \\
P18 & Consultant & Asia-Pacific \\
P19 & Data Scientist & South America \\
P20 & Data Scientist & Asia-Pacific \\
P21 & Data Scientist & Asia-Pacific \\
P22 & Developer & Asia-Pacific \\
\hline
\end{tabular}
\end{table}

\subsection{Dataset and Task}

The dataset needed to be substantial enough to allow researchers to move beyond obvious initial hypotheses, requiring both breadth and depth of analysis to generate interesting and useful insights. We selected the 2020 US Census dataset\footnote{https://www.kaggle.com/datasets/takumafujiwara/2020-census-data} for our study, which contains approximately 40,000 rows with 15 columns including age, workclass, education, marital status, occupation, and income. Participants were asked to formulate hypotheses related to \textbf{income inequality} with the following task description:

\begin{quote}
    ``We are seeking skilled data scientists and researchers to analyze the 2020 US Census data and identify novel insights into the factors contributing to the widening income gap in America. This competition aims to go beyond surface-level explanations and uncover nuanced, data-driven findings that can inform policy decisions and public discourse."
\end{quote}

\subsection{User Study Procedure}
\label{sec:study_procedure}

Our study procedure is outlined as follows:


\begin{enumerate}
        \item \textbf{Introduction} (5 minutes): We introduced the task, dataset, and system features to participants, followed by a guided example to ensure familiarity with the tool.
        
        \item \textbf{Pre-study Survey} (2 minutes): Participants answered questions about their current hypothesis formulation practices and use of AI tools in their workflow:
        \begin{itemize}
            \item Q1: How did you formulate hypothesis in your daily work?
            \item Q2: Do you use GenAI tools to formulate hypothesis? Which tools?
        \end{itemize}
        
        \item \textbf{Hypothesis Formulation Task} (15 minutes): Participants used our system to formulate hypotheses related to income inequality. We employed a think-aloud protocol during this phase.
        
        \item \textbf{Structured Interview} (8 minutes): We conducted in-depth interviews to gather qualitative data on participants' experiences with the following questions:
        \begin{itemize}
            \item Q1: Did the tool help you formulate hypothesis? How? (if not, why not?)
            \item Q2: Is the visualization (tree diagram) helpful for you for exploring the data? How does the visualization help you understand the dataset?
            \item Q3: What new information/function/ways of interacting with the data did you want for formulating hypothesis? Why?
            \item Q4: There is not much natural language interaction here, do you want some new functions like chatbot?
            \item Q5: After using this tool, do you want to use GenAI tools for hypothesis formulation in the future? Why?
    \end{itemize}
\end{enumerate}

\subsection{Data Collection and Analysis Methods}

As mentioned earlier in this section, we collected both study interaction data and interview data. The study interaction data included node-link diagrams created by participants in JSON format, user interactions such as node clicking, hypothesis generation, and visual hypothesis expansion, and video recordings of the study process obtained with participants' consent. For the interview data, we conducted and recorded user interviews, which were automatically transcribed and manually corrected.

Our analysis approach incorporated both quantitative and qualitative methods to examine the user study data. We employed three primary analysis methods:

\begin{enumerate}
    \item \textbf{Analysis of the Node-Link Diagram Structure}: We analyzed the node-link diagram structure created by participants. This involved extracting metrics of the diagram structures such as the maximum breadth and depth of each diagram and count of nodes by level.
    \item \textbf{Analysis of User Interactions}: We examined the user interaction data to understand how participants engaged with the tool. This included analyzing patterns of node clicking, hypothesis generation, and visual hypothesis expansion.
    \item \textbf{Thematic Analysis of Structured Interviews}: To understand participants' experiences and perceptions in their own words, we conducted a thematic analysis of the interview responses.
\end{enumerate}

\section{Results}
\label{sec:results}

\begin{figure*}
\includegraphics[width=1.\textwidth]{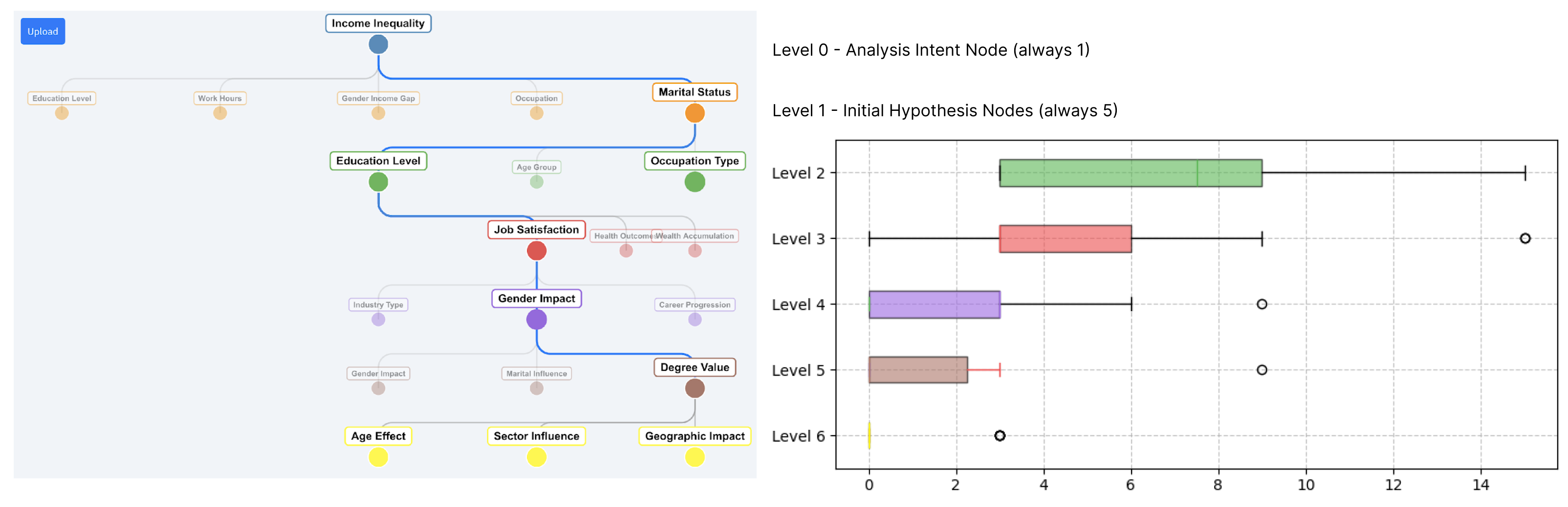}
  \caption{Count of generated nodes by level, starting from level 2 given that level 0 always has 1 analysis intent node and level 1 always has 5 initial hypothesis nodes. As the depth increases, the count of generated nodes per level progressively decreases, reflecting a more focused analysis at deeper levels of the tree structure.}
  \Description{}
  \label{fig:node-count-by-level}
\end{figure*}

\begin{figure*}
\includegraphics[width=0.8\textwidth]{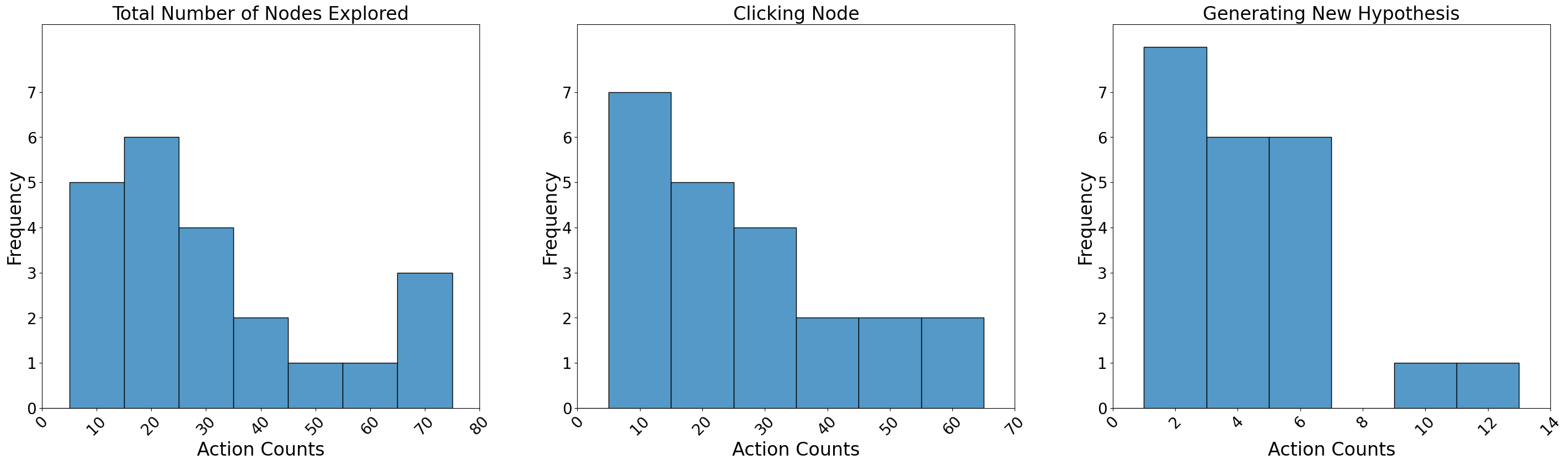}
  \caption{Counts of total number of nodes explored (number of nodes clicked + number of new hypothesis branch generated, a new node will be displayed when a new branch is generated), number of nodes clicked, number of new hypothesis branch generated.}
  \Description{}
  \label{fig:nodes-count}
\end{figure*}

\begin{table*}
\centering
\begin{tabular}{|p{0.02\textwidth}|p{0.16\textwidth}|p{0.65\textwidth}|p{0.05\textwidth}|}
\hline
\textbf{ID} & \textbf{Hypothesis} & \textbf{Description} & \textbf{Level}\\
\hline
P1 & Age Effect & There is a variance in income between married and non-married individuals within different age groups. & 3 \\
\hline
P2 & Occupation Type & There is a significant difference in income levels based on occupation types between married and unmarried individuals. & 2 \\
\hline
P2 & Gender Impact & There is a significant correlation between holding an advanced degree and higher incomes with the additional factor of gender, affecting married individuals differently compared to unmarried individuals. & 4 \\
\hline
P3 & Community Integration & There is a stronger positive correlation between Ivy League education and higher lifetime income for underrepresented racial groups in highly integrated communities. & 4 \\
\hline
P4 & Experience Amplifier & There is a \textbf{compounding effect} of work experience on income, which is amplified with each additional degree. & 2 \\
\hline
P5 & Investment Growth & Higher income earners exhibit a faster growth rate of their financial assets over time. & 5 \\
\hline
\end{tabular}
\caption{AI-generated hypotheses bookmarked by Participant 1-5.}
\label{tab:hypotheses}
\end{table*}

\subsection{How did users balance breadth and depth of exploration?}

To contextualize our analysis of the breadth and depth of hypothesis exploration by the participants, we describe overall quantitative trends in participants' interactions with the tool. We analyzed the node-link diagram structures generated during the study. On average, diagrams contained 21.82 nodes ($std = 8.78$), reached a maximum depth of 3.86 levels ($std = 1.32$), and had a maximum breadth of 8.09 ($std = 3.06$). Figure \ref{fig:node-count-by-level} illustrates the distribution of generated nodes across different levels. To account for the actual number of nodes explored by each participant, rather than just the number of nodes created, we tracked two actions: clicking existing nodes and generating new hypotheses. As illustrated in Figure \ref{fig:nodes-count}, participants clicked nodes an average of 31.41 times ($std = 17.17$) and generated new hypotheses 4.95 times ($std = 2.72$). In total, participants explored nodes an average of 36.36 times ($std = 18.89$).

To further illustrate the generated hypotheses and the hypothesis exploration and iteration process, we present examples of AI-generated hypotheses bookmarked by the first five participants in Table \ref{tab:hypotheses}. These examples demonstrate how the generated hypotheses went beyond simple pairwise correlation predictions, incorporated complex and nuanced ideas, such as the ``compounding effect."

Based on the interview results for Q5 measuring overall attitudes toward the system, the hypothesis exploration tool received positive feedback from 20 out of 22 participants, such as P12: ``I absolutely would use this, 100\%." One participant expressed selective interest, showing appreciation for the tool's data visualization features but remaining cautious about its utility for hypothesis formulation. Another participant cited a commitment to originality and privacy concerns: ``I'm trying to create something novel myself. I don't want to have to cite it or accidentally reveal one of my ideas."

\subsubsection{Structured Workflow and Focused Exploration}

While the diagram supported broad exploration, it also provided a structured framework that some users likened to ``guardrails". This structure helped focus users' thinking without overly constraining their creativity. A developer (P10) described this balance:

\begin{quote}
``\textbf{The tree diagram is like guardrails} I would say yes it does (help me), because it limits my imagination down to kind of selectable options. So it puts boundaries on my creativity [while] still allowing me to, like, follow different paths I might have, that I might come up with. So like guardrails I guess would probably be the best way to sum it up. \textbf{The guardrails it imposes focuses me}." (P10, Developer)
\end{quote}

The tree-like structure of the diagram also facilitated drilling down into specific hypotheses. For instance, P19 (a data scientist) described how they were able to use the representation to put ``hypothesis into \textbf{buckets}. So you know that you will have a hypothesis for location level, a hypothesis for the work hours. And at the end, I think that this advantage of that tree diagram could be that \textbf{you only focus on one topic and forget about the rest}." Similarly, P13 (a consultant) said, ``[the diagram] is really helpful in connecting from one hypothesis to another hypothesis until it achieves my goal, right? That is the best part of [the] tree diagram, that helped me to drive through each and every hypothesis at each level and \textbf{drill down to my goal}.''



Participants found the logical flow and visual organization of the diagram conducive to comprehending AI-generated multi-thread hypotheses:

\begin{quote}
``It just gives you a \textbf{logical flow} [..] like, which variables belong in which. It's like a visualization of where variables belong, where they're clustered. Yeah, it just means you go through it very quickly." (P12, UX researcher)
\end{quote}

\begin{quote}
``It allows you to figure out how things are connected. I thought \textbf{it was pretty easy to read and follow}. This helps me understand the data set by helping me understand what different hypothesis were connected and how [and] under [..] what topic." (P3, Global sales, Figure \ref{fig:teaser})
\end{quote}

\begin{quote}
``I mean its user interface is super easy to use. I mean you can click and explore. Uh, and I feel like \textbf{it's very easy to read} because the options are spread out enough. I mean, as the tree gets further down, it gets more specific. So I like that." (P5, Developer, Figure \ref{fig:user-journey})
\end{quote}

\subsubsection{Supporting Overview and Backtracking}

The ordered node-link diagram offered a hierarchically structured visualization of the entire hypothesis space. An advantage of this overview was its ability to facilitate backtracking. This feature empowered users to revisit and reconsider earlier hypotheses, enhancing the breadth of their exploration. As one consultant (P18) noted:

\begin{quote}
``This visualization tree, \textbf{it's like [an] organized hierarchy} [..]. It was a structured and where we can easily get to know like, if we miss it somewhere, for example, in the educational level option, if you miss somewhere then to \textbf{recreate} and to \textbf{reread} it. So that time, this hierarchy level tree structure visualization helps us to \textbf{check again}, \textbf{recheck and analyze it}." (P18, Consultant)
\end{quote}

The visual structure also helped users identify areas of focus within their analysis:

\begin{quote}
``It helps to determine \textbf{which area it is more focused}, like which area it is affecting, like for what is the main reason of in income inequality and education level, in which area of education level it's more affecting. What is the main cause of income inequality?" (P9, Developer)
\end{quote}

\subsection{How did users engage in non-linear exploration of hypotheses?}

\begin{figure*}
\includegraphics[width=1\textwidth]{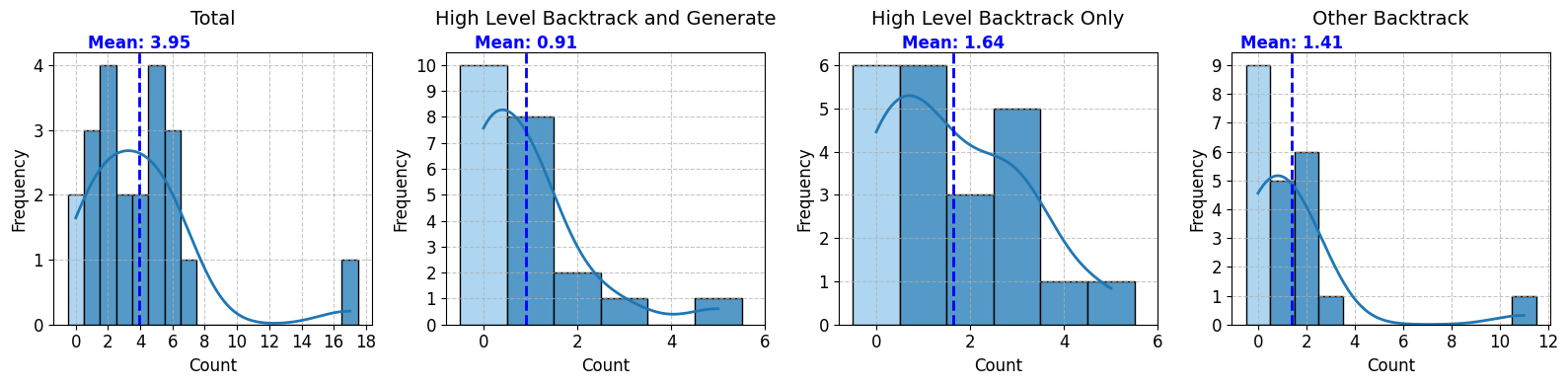}
  \caption{Histograms depicting backtrack behaviors: backtracking previously created branches. The data illustrates three types of backtrack behaviors: high level backtrack and generate, high level backtrack only, and other backtrack instances. The breakdown for each participant can be found in Table \ref{tab:backtrack} in Appendix \ref{appendix:C}.}
  \Description{}
  \label{fig:backtrack-hist}
\end{figure*}

Some participants exhibited non-linear exploration patterns, exemplified by the backtracking behaviors of P3 (Figure \ref{fig:teaser} and Figure \ref{fig:P3} in Appendix \ref{appendix:B}). P3's process, which included 17 instances of backtracking, prompted a further analysis of this behavior across all participants. Figure \ref{fig:backtrack-hist} illustrates various types of backtracking behaviors, including \textit{high level backtrack and generate} (revisit a hypothesis node on a higher level and not on the trajectory from the current node to the root node, and generate new branch from the node), \textit{high level backtrack only} (only revisit a hypothesis node on a high level and not on the current trajectory) and \textit{other backtrack} instances (revisit a hypothesis node but not on a higher level).

\subsubsection{Rapidly Prototyping Alternative Hypotheses to Avoid Fixation}

The node-link diagram enabled users to generate and explore multiple hypotheses simultaneously. This multi-threaded prototyping approach was particularly appreciated by participants for its ability to broaden their analytical perspective. As one UX researcher (P12) noted:

\begin{quote}
    ``What I thought [..] was really cool is every time you've done your piece of analysis, as you go to your supervisor and you say, `Look, this is what I found.' 'Ohh that's really interesting. I wonder if you looked at it this way', but you never have time to do that... So from that point of view, I think \textbf{it's really good for those additional hypothesis} that you might want to go in and test." (P12, UX researcher)
\end{quote}

This sentiment was echoed by other participants, who valued the ability to explore multiple options and dimensions:

\begin{quote}
    ``When you propose a new hypothesis, \textbf{it does not give you only one option}. It gives you more than one, so you can start like exploring other branches and so that was really helpful." (P19, Data scientist)
\end{quote}

\begin{quote}
    ``It gives me the options instead of me thinking of it. It gives me the \textbf{multiple options} based on my hypothesis. What I am thinking of right? So that dimension it is giving it for the next level of hypothesis. I want to create it, so that is what the tool is helping us. That is the best part of it. So once I finalize job rule right, then the successive hypothesis, when it creates it gives 3 dimensions right? So one is \textbf{different dimensions}, that is what it is helping us." (P13, Consultant)
\end{quote}

Further, participants described benefits of avoiding fixation and bias of having these hypotheses in front of them in the system. As a concrete example, P4 (who worked in Business Technology) described how different the hypotheses they explored here were from what they might have started with:

\begin{quote}
    ``There was probably \textbf{prior biases}...if you were to ask me like without having seen any of this tool, what do you think correlates to higher income? I probably could have told you years of experience and like number of the degrees." (P4, Business technology)
\end{quote}

P12 expanded on how this benefit might stem from the hypothesis space being "there in front of you", inviting exploration: 

\begin{quote}
    ``And I think as well, like researchers tend to get, tend to get \textbf{dug in with particular variables}, say like if you're if genders on your mind, they'll think of gender, but maybe it won't cross their mind to look at marital status. But when it's \textbf{there in front of you}, you're gonna `Ohh yeah, why don't I think? Let's have a look at that as well'." (P12, UX researcher)
\end{quote}

P3 (who worked in a Global Sales role) described how this potential benefit of mitigating prior biases might be a more generally useful thing for analysts:

\begin{quote}
    ``I think it is a very useful tool that would help kind of \textbf{take away bias} to a certain extent. I feel like we have a lot of bias or too much personal views that can kind of formulate hypothesis." (P3, Global sales, Figure \ref{fig:teaser})
\end{quote}


\subsubsection{Multi-thread vs Single-thread? Perspectives on Including Chatbot in the Interface}

In Q4 of our semi-structured interviews, participants shared their personal experiences and reflections on the idea of incorporating a chatbot into the interface. A number of participants explained that the current multi-threaded visualization felt like a natural extension of their workflow of hypothesis exploration and analysis. For example, P8 conveyed how the detailed, structured flow of information resonated with their daily practices:

\begin{quote}
``\textbf{Chatbot is only end-user oriented}. We (data scientists) need something \textbf{intermediary}... when I go to the manager, he needs the insight only. He doesn't want that full narrative." (P8, Data scientist)
\end{quote}

This reflection illustrates how P8 found the multi-threaded visualization helpful for uncovering layered insights--a richness that might be compromised by a more streamlined chatbot interaction. Conversely, their managers often need only concise, final insights, which could potentially be fulfilled by a chatbot interface. Other participants expressed interests for the inclusion of a chatbot that might bring a conversational, debate-like dynamic to their exploration process and enhance the depth of exploration:

\begin{quote}
``I'd like a chatbot. I think it would be nice to be able to talk to the hypothesis as if you were talking to \textbf{someone defending that hypothesis}, because I [..] feel like that's what people do." (P2, Developer, Figure \ref{fig:node-count-by-level})
\end{quote}

\begin{quote}
``Yeah, actually chatting about a hypothesis. Like let's say I get to hear and I wanted to try discussing it. \textbf{Like a debater} at this point that I could then [be] either for or against." (P10, Developer)
\end{quote}

\subsection{How did users engage with relevant data while exploring hypotheses?}

\begin{figure*}
\includegraphics[width=1\textwidth]{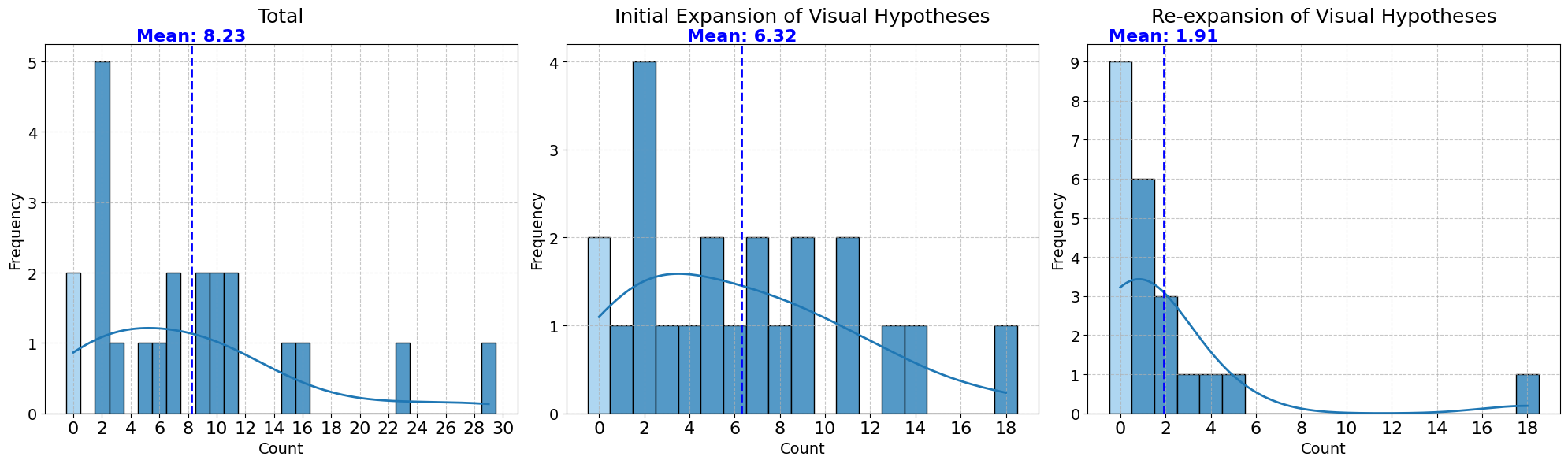}
  \caption{Histograms depicting user engagement with visual hypotheses. The data illustrates two types of interactions: initial expansion (opening a visual hypothesis for the first time) and re-expansion (reopening a previously viewed visual hypothesis). Light blue bars represent participants who did not (re-)expand any visual hypothesis. The breakdown for each participant can be found in Table \ref{tab:engagement} in Appendix \ref{appendix:B}.}
  \Description{}
  \label{fig:engage-hist}
\end{figure*}

We examined user engagement with AI-generated visualizations corresponding to each hypothesis, with Figure \ref{fig:engage-hist} depicting the frequency of initial expansions and re-expansions of these visual hypotheses.

\subsubsection{Visual Hypotheses Enable Judgment of Hypothesis Grounding in Data}
Participants shared that the visual hypotheses generated by the system suggested whether there were evidences of their ideas or not. This visual representation helped transform abstract thoughts into data-backed concepts. P4 and P8 described this benefit:

\begin{quote}
``I will say that the visualization of the census data was helpful in \textbf{confirming} the potential income based on the higher degrees. So it was \textbf{affirming} of that. I'm further also just the visual on the working hours and how scattered that was also \textbf{affirmed}." (P4, Business technology)
\end{quote}

\begin{quote}
``It showed me data that was \textbf{reaffirming} to my hypothesis...Like it would just be like an idea of my head that would have had no backing. The fact that I saw it with data is like OK, now this has more credibility. I'm more willing to follow it than I would have if I had just thought of it." (P8, Data scientist)
\end{quote}

Participants noted and appreciated the direct nature of visual hypotheses. As one researcher (P1) observed, ``I guess most useful part is the figure like the visualization...hypothesis, it's kind of indirect." This sentiment was echoed by a data scientist (P19):

\begin{quote}
``I think it's really really useful because at the end the line charts, the line graph, \textbf{it's really useful to have like a graph that shows or that is directly associated with the hypothesis} that you are seeing...I think having the charts in here, it's really really useful, to really understand what is happening, to be sure that the hypothesis that you are proposing is tied to a graph"
\end{quote}

\subsubsection{Blending Hypotheses and Charts}
Participants mentioned that an advantage of the multimodality was the seamless integration of hypotheses and visualizations. This feature reduced the effort associated with context switching between hypothesis formulation and chart creation. A data scientist (P8) enthusiastically described how this integration enhanced their workflow:

\begin{quote}
``It helps me very much and then really we love it. Really. I love it. I can say definitely it helps. The chart and the narrative. And then I get immediately and instantly, which can be transferred visualization I can say because usually we struggle with the visualization and the content when GenAI gives both. It's really easier for us."
\end{quote}

Another data scientist (P19) shared how efficiency was gained from this blended approach:

\begin{quote}
``I will say that it's really easy to get hypothesis and get like charts that support the hypothesis really quick. So that's an easy way to maybe start exploring data and see what the data has, and go through there, and start like going a little bit more specific."
\end{quote}





\section{Discussion}

\subsection{Summary of Results}

In this paper, we tested the design of a shared representation for GenAI-assisted hypothesis exploration with a design probe (n=22). The majority of our study participants were enthusiastic and excited about this tool and would use it in the future for hypotheses exploration. The proposed ordered node-link diagram serves as a potential shared representation, providing structure while allowing flexibility in exploration. Participants likened it to ``guardrails" that focused their thinking without constraining creativity. The diagram facilitated overviews of the hypothesis space and supported frequent backtracking, enabling broader exploration. On average, users created 21.82 hypotheses (SD = 8.78) and reached a maximum depth of 3.86 levels (SD = 1.32), demonstrating exploration patterns including high-level backtracking and branch regeneration. The multimodal information hints, particularly visual hypotheses, helped users judge whether a hypothesis was worth pursuing based on how it might be grounded in the data. The integration of hypotheses and charts reduced context switching effort, while supporting text provided valuable context. These features collectively supported breadth-oriented exploration and depth-oriented iteration in hypothesis formulation.

\subsection{Limitation and Future Work}

\subsubsection{Potential System Improvements}

The ordered node-link diagram, while generally well-received, had room for improvement according to participant feedback in the interview. Suggestions included enhancing user control over hypothesis exploration, allowing direct manipulation of dataset columns, connecting relevant hypotheses, and showing branched nodes with different user inputs. 

\re{Another limitation arises from the scalability of node-link diagrams: as more hypotheses are explored, the diagram's complexity increases rapidly. In the short time period of our study, participants explored on the order of 20-30 hypothesis nodes, on average: it is possible that, accounting for variations in screen real estate across devices, that this is close to the practical upper bound for what can be visually managed. It is also possible, however, that in the context of larger data analysis workflows, the scale of exploration might not extend much more beyond tens of hypothesis nodes. 
Future work should systematically evaluate hypothesis exploration using node-link diagrams in more ecologically valid settings, such as with problems of varying complexity, or over longer time periods, to better understand these scalability limitations, and explore how augmenting the node link diagram with previously studied techniques, such as dynamic rescaling and filtering/aggregation of branches \cite{plaisantSpaceTreeSupportingExploration2002} or partial overviews \cite{leeTreePlusInteractiveExploration2006}, might overcome these limitations}.

Participants also provided insights on enhancing the multimodal information hints. They proposed including sources for preliminary data visualization and related work to improve output explainability. Another suggestion was to convert the generated information into preliminary presentation formats like slides, which could facilitate easier review by data science teams and make the information more accessible.

\re{Finally, we envision our system functioning as a component in data analysis workflows, to be used during conceptual hypothesis exploration, rather than replacing entire existing workflows, such as EDA. How our prototype system might interact with other components of data analysis workflows remains an open question. Perhaps more specific EDA visualizations might enhance the information hints in our interface, and selected hypotheses from our system could be "exported" to statistical model planning systems like rTisane for further refinement \cite{junRTisaneExternalizingConceptual2024}.}

\subsubsection{Study Limitation and Future Work}

The study has \re{three} main limitations that present opportunities for future research. \re{First, our research questions and study design here focused on qualitative understanding of the user experience; thus, our data cannot speak to quantitative or comparative effects of our system on the quality or creativity of hypothesis exploration}. To address this, future work should incorporate \re{direct comparisons to baseline or other systems in appropriate between or within-subjects studies; these studies could compare process-based metrics (e.g., number/diversity/depth of hypotheses explored) as well outcome-based metrics (probability of insights, overall quality/novelty of hypotheses)}.

Additionally, the current task setting is confined to a laboratory environment, using a sample 2020 US census dataset and a predetermined analysis intent (``investigate the data from the perspective of income inequality") that may not fully reflect participants' real-world data analysis tasks. To enhance ecological validity and broaden the applicability of findings, future research should allow participants to bring their own datasets and define their own analysis tasks or intents. This approach could lead to an ``in the wild" deployment, where the system is tested in authentic work environments over extended periods. Such real-world setting would provide deeper insights into how the hypothesis exploration tool integrates into actual data analysis workflows and its long-term impact on conceptual reasoning processes.

\re{Third, our system builds on prior work that benchmarks the quality, novelty, and other dimensions of AI-generated hypotheses from LLMs comparable to those we used \cite{siCanLLMsGenerate2024, baekResearchAgentIterativeResearch2025}. Therefore, we assumed a baseline level of quality and did not independently conduct a comprehensive evaluation of the generated hypotheses. While Table \ref{tab:hypotheses} presents hypothesis examples, we were not yet in a position to make claims about the overall quality of the generated content. Future work could evaluate these hypotheses across multiple dimensions, including feasibility, potential impact, and novelty.}

\begin{figure*}
\includegraphics[width=1\textwidth]{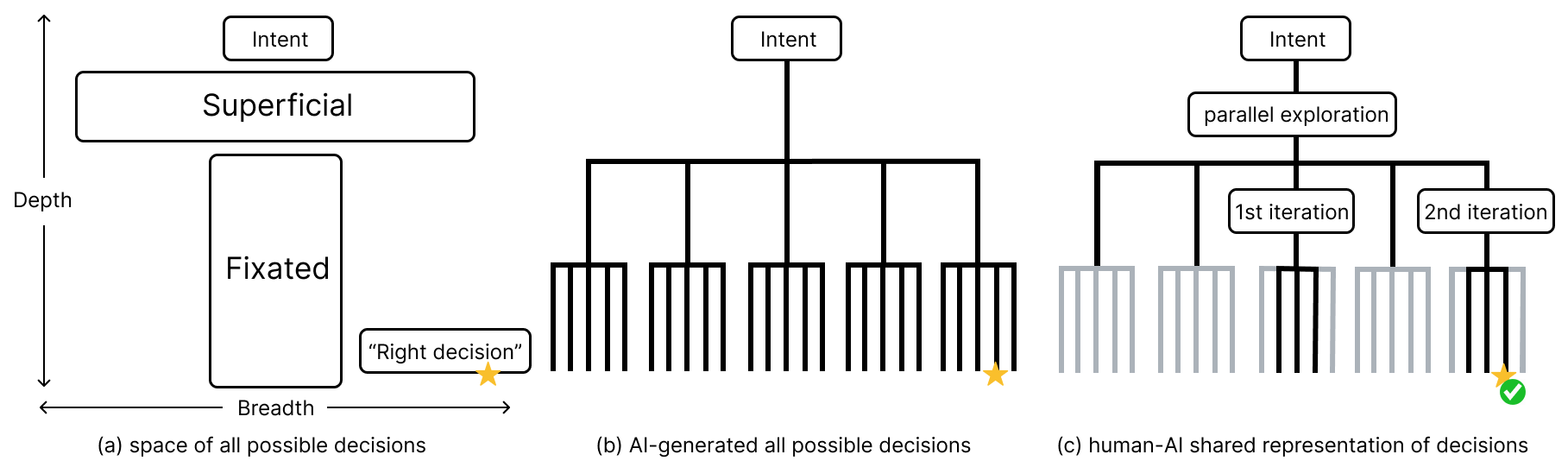}
  \caption{Conceptual modeling of decision making space in a 2D space, similar to Chan et al. \cite{chanFormulatingFixatingEffects2024}. (a) (b) (c) represent the same space. (a) represents of the space of possible decisions, showing areas of superficiality and fixation relative to human intent, with the ``right decision" represented with a star potentially lying outside these areas. (b) AI-generated multi-thread exploration of all possible decisions, illustrating maximum breadth and depth. (c) A human-AI shared representation of decisions, balancing parallel exploration with iteration to efficiently navigate the decision making space.}
  \Description{}
  \label{fig:model}
\end{figure*}

\subsection{Beyond Hypothesis Exploration: Guiding Decision Making and Coordinating Complex Workflows}

\subsubsection{Conceptual Guardrails for Decision Making: Balancing Exploration Depth and Breadth to Overcome Superficiality and Fixation}

Our findings demonstrate that the interactive node-link diagram can serve as an effective shared representation for hypothesis exploration, striking a balance between guidance and discovery. This concept extends beyond hypothesis exploration, suggesting broader implications for decision-making processes supported by ordered node-link diagrams with information scents.

Making context-appropriate decisions is challenging due to two primary factors: superficiality (lack of depth) \cite{kahneman2011thinking} and fixation (lack of breadth) \cite{chanFormulatingFixatingEffects2024,janssonDesignFixation1991}. We propose modeling the space of possible decisions in a two-dimensional model, inspired by Chan et al. \cite{chanFormulatingFixatingEffects2024}, as illustrated in Figure \ref{fig:model}(a). In this model, proximity to human initial intent potentially indicates superficiality, while perpendicular directions may represent fixation. The ``right decision", depicted as a star, often lies outside these areas. To identify this ``right decision," we need to employ both iterative processes to increase depth and overcome superficiality, and parallel exploration to expand breadth and mitigate fixation.

Given the constraints of limited time budgets, there exists an inherent trade-off between breadth-oriented exploration and depth-oriented iteration. Generative AI offers a promising solution to this dilemma by leveraging its extensive world knowledge and generative capabilities to produce potential correct pathways efficiently. This AI-driven approach can complement human expertise, allowing domain specialists to apply their knowledge and decision-making skills in selecting and refining analytical pathways. To facilitate this human-AI collaboration effectively, a carefully designed shared representation is essential, one that harmonizes parallel exploration with iterative refinement.

While leveraging GenAI to generate an exhaustive set of possibilities could theoretically maximize both exploration and iteration, as depicted in Figure \ref{fig:model}(b), this approach risks overwhelming human cognitive capacities with an excess of information. Our research suggests a more nuanced solution: the strategic use of information hints. These hints serve as cognitive scaffolds, guiding users to pause or terminate unpromising trajectories, thereby streamlining the parallel exploration and iteration process, as illustrated in Figure \ref{fig:model}(c). The node-link diagram interface with AI-generated hints, offers a promising framework for enhancing decision-making processes across various domains.

Additionally, this focus on high-level ideation and decision-making within the shared representation holds potential for collecting domain experts' thinking processes. With ethical and secure handling, the collected data can be employed for reinforcement learning with human factors (RLHF) to enhance AI models' hypothesis generation capabilities. For instance, AI could generate three hypotheses based on intent, and the expert's selection of one to branch further could serve as the winning response for RLHF, while the remaining two could be used as losing responses. This iterative process offers a pathway to continually refine AI's ability to generate expert-level hypotheses.

\subsubsection{Coordinating Subtasks in Complex Task Workflows}

The findings from our study on hypothesis exploration have broader implications for coordinating and managing complex tasks that involve multiple interconnected subtasks. While AI systems have shown strong potential in handling isolated tasks, such as generating news headlines \cite{dingHarnessingPowerLLMs2023}, research ideas \cite{siCanLLMsGenerate2024}, or data visualization \cite{dibiaLIDAToolAutomatic2023}, they still face limitations when dealing with intricate, multi-faceted workflows \cite{xiaRedesigningInformationSpace2024,dingAdvancingGUIGenerative2024}. As an example of complex workflows, data storytelling video production involves multiple components including data analysis, story writing, prototyping, image editing, animation creation, and sound effect generation, each of which can influence and necessitate iterations in other subtasks \cite{xiaRedesigningInformationSpace2024}.

The interactive diagram we developed for hypothesis exploration could be adapted as a shared representation to visualize and manage the inter-dependencies between individual pieces in complex workflows. This aligns with existing work on AI-powered data analytics workflows that utilize similar shared representations \cite{xieWaitGPTMonitoringSteering2024}. The overview and backtracking capabilities demonstrated in our system could be valuable in complex task management. They would allow humans to maintain a holistic view of the project while also being able to drill down into specific areas or revert to previous stages when necessary. This flexibility is important in workflows where changes in one area can have cascading effects on others. The finding that the system supports rapid prototyping of alternatives could be beneficial in enabling users to quickly explore different approaches or solutions. Our design of information hint panel could also be expanded to serve as a coordination tool between subtasks to make informed decisions and ensure alignment across different AI-powered pieces in complex tasks.


\section{Conclusion}


This paper introduced the system design and empirical results from a user study of a novel interactive diagram that serves as a shared representation for AI-assisted hypothesis exploration in data analysis, addressing the need for intelligent support in conceptual tasks. We proposed and implemented a prototype of an ordered node-link diagram augmented with AI-generated visual and textual information hints. Our design probe (n=22) provides strong support for the effectiveness and usefulness of this ordered node-link diagram, with participants generating an average of 21.82 hypotheses. Our detailed usage analysis also indicated that the diagram was effective at both helping participants explore both breadth and depth of the hypotheses space. For many, the interactive diagram functioned as ``guardrails" for hypothesis exploration, providing structured workflows and overviews, and efficient backtracking capabilities. AI-generated visual and textual information hints helped users transform abstract ideas into data-backed concepts while reducing context switching effort. While we observed how our approach can potentially enhance the breadth and depth of human-AI collaborative data analysis, we see a lot of promise in this technique for managing conceptual tasks such as providing conceptual guardrails for more general decision-making in complex AI-assisted workflows that go beyond simple interactions with large language models.



\bibliographystyle{ACM-Reference-Format}
\bibliography{sample-base}


\begin{thebibliography}{50}


\ifx \showCODEN    \undefined \def \showCODEN     #1{\unskip}     \fi
\ifx \showDOI      \undefined \def \showDOI       #1{#1}\fi
\ifx \showISBNx    \undefined \def \showISBNx     #1{\unskip}     \fi
\ifx \showISBNxiii \undefined \def \showISBNxiii  #1{\unskip}     \fi
\ifx \showISSN     \undefined \def \showISSN      #1{\unskip}     \fi
\ifx \showLCCN     \undefined \def \showLCCN      #1{\unskip}     \fi
\ifx \shownote     \undefined \def \shownote      #1{#1}          \fi
\ifx \showarticletitle \undefined \def \showarticletitle #1{#1}   \fi
\ifx \showURL      \undefined \def \showURL       {\relax}        \fi
\providecommand\bibfield[2]{#2}
\providecommand\bibinfo[2]{#2}
\providecommand\natexlab[1]{#1}
\providecommand\showeprint[2][]{arXiv:#2}

\bibitem[\protect\citeauthoryear{Ackerman, Dachtera, Pipek, and Wulf}{Ackerman et~al\mbox{.}}{2013}]%
        {ackermanSharingKnowledgeExpertise2013}
\bibfield{author}{\bibinfo{person}{Mark~S. Ackerman}, \bibinfo{person}{Juri Dachtera}, \bibinfo{person}{Volkmar Pipek}, {and} \bibinfo{person}{Volker Wulf}.} \bibinfo{year}{2013}\natexlab{}.
\newblock \showarticletitle{Sharing {Knowledge} and {Expertise}: {The} {CSCW} {View} of {Knowledge} {Management}}.
\newblock \bibinfo{journal}{\emph{Computer Supported Cooperative Work (CSCW)}} \bibinfo{volume}{22}, \bibinfo{number}{4-6} (\bibinfo{date}{Aug.} \bibinfo{year}{2013}), \bibinfo{pages}{531--573}.
\newblock
\showISSN{0925-9724, 1573-7551}
\urldef\tempurl%
\url{https://doi.org/10.1007/s10606-013-9192-8}
\showDOI{\tempurl}


\bibitem[\protect\citeauthoryear{{Alzheimer's Research UK}}{{Alzheimer's Research UK}}{2022}]%
        {AlzheimersResearchUK2022}
\bibfield{author}{\bibinfo{person}{{Alzheimer's Research UK}}.} \bibinfo{year}{2022}\natexlab{}.
\newblock \bibinfo{title}{Explaining the amyloid research study controversy}.
\newblock
\newblock
\urldef\tempurl%
\url{https://www.alzheimers.org.uk/for-researchers/explaining-amyloid-research-study-controversy}
\showURL{%
\tempurl}
\newblock
\shownote{Accessed: [Insert access date here].}


\bibitem[\protect\citeauthoryear{Baek, Jauhar, Cucerzan, and Hwang}{Baek et~al\mbox{.}}{2025}]%
        {baekResearchAgentIterativeResearch2025}
\bibfield{author}{\bibinfo{person}{Jinheon Baek}, \bibinfo{person}{Sujay~Kumar Jauhar}, \bibinfo{person}{Silviu Cucerzan}, {and} \bibinfo{person}{Sung~Ju Hwang}.} \bibinfo{year}{2025}\natexlab{}.
\newblock \bibinfo{title}{{ResearchAgent}: {Iterative} {Research} {Idea} {Generation} over {Scientific} {Literature} with {Large} {Language} {Models}}.
\newblock
\newblock
\urldef\tempurl%
\url{https://doi.org/10.48550/arXiv.2404.07738}
\showDOI{\tempurl}
\newblock
\shownote{arXiv:2404.07738 [cs].}


\bibitem[\protect\citeauthoryear{Bao, Li, Flores, Correll, and Battle}{Bao et~al\mbox{.}}{2022}]%
        {baoRecommendationsVisualizationRecommendations2022}
\bibfield{author}{\bibinfo{person}{Calvin Bao}, \bibinfo{person}{Siyao Li}, \bibinfo{person}{Sarah Flores}, \bibinfo{person}{Michael Correll}, {and} \bibinfo{person}{Leilani Battle}.} \bibinfo{year}{2022}\natexlab{}.
\newblock \bibinfo{title}{Recommendations for {Visualization} {Recommendations}: {Exploring} {Preferences} and {Priorities} in {Public} {Health}}.
\newblock
\newblock
\urldef\tempurl%
\url{http://arxiv.org/abs/2202.01335}
\showURL{%
\tempurl}
\newblock
\shownote{arXiv:2202.01335 [cs].}


\bibitem[\protect\citeauthoryear{Cao, E, Chen, and Xia}{Cao et~al\mbox{.}}{2023}]%
        {caoDataParticlesBlockbasedLanguageoriented2023}
\bibfield{author}{\bibinfo{person}{Yining Cao}, \bibinfo{person}{Jane~L E}, \bibinfo{person}{Zhutian Chen}, {and} \bibinfo{person}{Haijun Xia}.} \bibinfo{year}{2023}\natexlab{}.
\newblock \showarticletitle{{DataParticles}: {Block}-based and {Language}-oriented {Authoring} of {Animated} {Unit} {Visualizations}}. In \bibinfo{booktitle}{\emph{Proceedings of the 2023 {CHI} {Conference} on {Human} {Factors} in {Computing} {Systems}}}, Vol.~\bibinfo{volume}{23}. \bibinfo{publisher}{ACM}, \bibinfo{address}{Hamburg Germany}, \bibinfo{pages}{1--15}.
\newblock
\urldef\tempurl%
\url{https://doi.org/10.1145/3544548.3581472}
\showDOI{\tempurl}


\bibitem[\protect\citeauthoryear{Cetina}{Cetina}{1999}]%
        {cetinaEpistemicCulturesHow1999}
\bibfield{author}{\bibinfo{person}{Karin~Knorr Cetina}.} \bibinfo{year}{1999}\natexlab{}.
\newblock \bibinfo{booktitle}{\emph{Epistemic {Cultures}: {How} the {Sciences} {Make} {Knowledge}}}.
\newblock \bibinfo{publisher}{Harvard University Press}, \bibinfo{address}{Cambridge, Mass}.
\newblock
\showISBNx{978-0-674-25894-5}


\bibitem[\protect\citeauthoryear{Chan, Ding, Kamrah, and Fuge}{Chan et~al\mbox{.}}{2024}]%
        {chanFormulatingFixatingEffects2024}
\bibfield{author}{\bibinfo{person}{Joel Chan}, \bibinfo{person}{Zijian Ding}, \bibinfo{person}{Eesh Kamrah}, {and} \bibinfo{person}{Mark Fuge}.} \bibinfo{year}{2024}\natexlab{}.
\newblock \bibinfo{title}{Formulating or {Fixating}: {Effects} of {Examples} on {Problem} {Solving} {Vary} as a {Function} of {Example} {Presentation} {Interface} {Design}}.
\newblock
\newblock
\urldef\tempurl%
\url{https://doi.org/10.48550/arXiv.2401.11022}
\showDOI{\tempurl}
\newblock
\shownote{arXiv:2401.11022 [cs].}


\bibitem[\protect\citeauthoryear{Chen and Xia}{Chen and Xia}{2022}]%
        {chenCrossDataLeveragingTextData2022}
\bibfield{author}{\bibinfo{person}{Zhutian Chen} {and} \bibinfo{person}{Haijun Xia}.} \bibinfo{year}{2022}\natexlab{}.
\newblock \showarticletitle{{CrossData}: {Leveraging} {Text}-{Data} {Connections} for {Authoring} {Data} {Documents}}. In \bibinfo{booktitle}{\emph{{CHI} {Conference} on {Human} {Factors} in {Computing} {Systems}}}. \bibinfo{publisher}{ACM}, \bibinfo{address}{New Orleans LA USA}, \bibinfo{pages}{1--15}.
\newblock
\showISBNx{978-1-4503-9157-3}
\urldef\tempurl%
\url{https://doi.org/10.1145/3491102.3517485}
\showDOI{\tempurl}


\bibitem[\protect\citeauthoryear{Chen, Xiong, Yao, and Glassman}{Chen et~al\mbox{.}}{2024}]%
        {zhu-tianSketchThenGenerate2024}
\bibfield{author}{\bibinfo{person}{Zhu-Tian Chen}, \bibinfo{person}{Zeyu Xiong}, \bibinfo{person}{Xiaoshuo Yao}, {and} \bibinfo{person}{Elena Glassman}.} \bibinfo{year}{2024}\natexlab{}.
\newblock \bibinfo{title}{Sketch {Then} {Generate}: {Providing} {Incremental} {User} {Feedback} and {Guiding} {LLM} {Code} {Generation} through {Language}-{Oriented} {Code} {Sketches}}.
\newblock
\newblock
\urldef\tempurl%
\url{http://arxiv.org/abs/2405.03998}
\showURL{%
\tempurl}
\newblock
\shownote{arXiv:2405.03998 [cs].}


\bibitem[\protect\citeauthoryear{Dibia}{Dibia}{2023}]%
        {dibiaLIDAToolAutomatic2023}
\bibfield{author}{\bibinfo{person}{Victor Dibia}.} \bibinfo{year}{2023}\natexlab{}.
\newblock \bibinfo{title}{{LIDA}: {A} {Tool} for {Automatic} {Generation} of {Grammar}-{Agnostic} {Visualizations} and {Infographics} using {Large} {Language} {Models}}.
\newblock
\newblock
\urldef\tempurl%
\url{http://arxiv.org/abs/2303.02927}
\showURL{%
\tempurl}
\newblock
\shownote{arXiv:2303.02927 [cs].}


\bibitem[\protect\citeauthoryear{Ding}{Ding}{2024a}]%
        {dingAdvancingGUIGenerative2024}
\bibfield{author}{\bibinfo{person}{Zijian Ding}.} \bibinfo{year}{2024}\natexlab{a}.
\newblock \showarticletitle{Advancing {GUI} for {Generative} {AI}: {Charting} the {Design} {Space} of {Human}-{AI} {Interactions} through {Task} {Creativity} and {Complexity}}. In \bibinfo{booktitle}{\emph{Companion {Proceedings} of the 29th {International} {Conference} on {Intelligent} {User} {Interfaces}}}. \bibinfo{publisher}{ACM}, \bibinfo{address}{Greenville SC USA}, \bibinfo{pages}{140--143}.
\newblock
\showISBNx{9798400705090}
\urldef\tempurl%
\url{https://doi.org/10.1145/3640544.3645241}
\showDOI{\tempurl}


\bibitem[\protect\citeauthoryear{Ding}{Ding}{2024b}]%
        {dingIntentbasedUserInterfaces2024}
\bibfield{author}{\bibinfo{person}{Zijian Ding}.} \bibinfo{year}{2024}\natexlab{b}.
\newblock \bibinfo{title}{Towards {Intent}-based {User} {Interfaces}: {Charting} the {Design} {Space} of {Intent}-{AI} {Interactions} {Across} {Task} {Types}}.
\newblock
\newblock
\urldef\tempurl%
\url{http://arxiv.org/abs/2404.18196}
\showURL{%
\tempurl}
\newblock
\shownote{arXiv:2404.18196 [cs].}


\bibitem[\protect\citeauthoryear{Ding and Chan}{Ding and Chan}{2024}]%
        {dingIntelligentCanvasEnabling2024}
\bibfield{author}{\bibinfo{person}{Zijian Ding} {and} \bibinfo{person}{Joel Chan}.} \bibinfo{year}{2024}\natexlab{}.
\newblock \bibinfo{title}{Intelligent {Canvas}: {Enabling} {Design}-{Like} {Exploratory} {Visual} {Data} {Analysis} with {Generative} {AI} through {Rapid} {Prototyping}, {Iteration} and {Curation}}.
\newblock
\newblock
\urldef\tempurl%
\url{https://doi.org/10.48550/arXiv.2402.08812}
\showDOI{\tempurl}
\newblock
\shownote{arXiv:2402.08812 [cs].}


\bibitem[\protect\citeauthoryear{Ding, Smith-Renner, Zhang, Tetreault, and Jaimes}{Ding et~al\mbox{.}}{2023}]%
        {dingHarnessingPowerLLMs2023}
\bibfield{author}{\bibinfo{person}{Zijian Ding}, \bibinfo{person}{Alison Smith-Renner}, \bibinfo{person}{Wenjuan Zhang}, \bibinfo{person}{Joel~R. Tetreault}, {and} \bibinfo{person}{Alejandro Jaimes}.} \bibinfo{year}{2023}\natexlab{}.
\newblock \bibinfo{title}{Harnessing the {Power} of {LLMs}: {Evaluating} {Human}-{AI} {Text} {Co}-{Creation} through the {Lens} of {News} {Headline} {Generation}}.
\newblock
\newblock
\urldef\tempurl%
\url{http://arxiv.org/abs/2310.10706}
\showURL{%
\tempurl}
\newblock
\shownote{arXiv:2310.10706 [cs].}


\bibitem[\protect\citeauthoryear{Dragicevic, Jansen, Sarma, Kay, and Chevalier}{Dragicevic et~al\mbox{.}}{2019}]%
        {dragicevicIncreasingTransparencyResearch2019a}
\bibfield{author}{\bibinfo{person}{Pierre Dragicevic}, \bibinfo{person}{Yvonne Jansen}, \bibinfo{person}{Abhraneel Sarma}, \bibinfo{person}{Matthew Kay}, {and} \bibinfo{person}{Fanny Chevalier}.} \bibinfo{year}{2019}\natexlab{}.
\newblock \showarticletitle{Increasing the {Transparency} of {Research} {Papers} with {Explorable} {Multiverse} {Analyses}}. In \bibinfo{booktitle}{\emph{Proceedings of the 2019 {CHI} {Conference} on {Human} {Factors} in {Computing} {Systems}}}. \bibinfo{publisher}{ACM}, \bibinfo{address}{Glasgow Scotland Uk}, \bibinfo{pages}{1--15}.
\newblock
\showISBNx{978-1-4503-5970-2}
\urldef\tempurl%
\url{https://doi.org/10.1145/3290605.3300295}
\showDOI{\tempurl}


\bibitem[\protect\citeauthoryear{Gu, Grunde-McLaughlin, McNutt, Heer, and Althoff}{Gu et~al\mbox{.}}{2024}]%
        {guHowDataAnalysts2024}
\bibfield{author}{\bibinfo{person}{Ken Gu}, \bibinfo{person}{Madeleine Grunde-McLaughlin}, \bibinfo{person}{Andrew McNutt}, \bibinfo{person}{Jeffrey Heer}, {and} \bibinfo{person}{Tim Althoff}.} \bibinfo{year}{2024}\natexlab{}.
\newblock \showarticletitle{How {Do} {Data} {Analysts} {Respond} to {AI} {Assistance}? {A} {Wizard}-of-{Oz} {Study}}. In \bibinfo{booktitle}{\emph{Proceedings of the {CHI} {Conference} on {Human} {Factors} in {Computing} {Systems}}}. \bibinfo{publisher}{ACM}, \bibinfo{address}{Honolulu HI USA}, \bibinfo{pages}{1--22}.
\newblock
\showISBNx{9798400703300}
\urldef\tempurl%
\url{https://doi.org/10.1145/3613904.3641891}
\showDOI{\tempurl}


\bibitem[\protect\citeauthoryear{Gu, Jun, and Althoff}{Gu et~al\mbox{.}}{2023}]%
        {guUnderstandingSupportingDebugging2023}
\bibfield{author}{\bibinfo{person}{Ken Gu}, \bibinfo{person}{Eunice Jun}, {and} \bibinfo{person}{Tim Althoff}.} \bibinfo{year}{2023}\natexlab{}.
\newblock \bibinfo{title}{Understanding and {Supporting} {Debugging} {Workflows} in {Multiverse} {Analysis}}.
\newblock
\newblock
\urldef\tempurl%
\url{http://arxiv.org/abs/2210.03804}
\showURL{%
\tempurl}
\newblock
\shownote{arXiv:2210.03804 [cs].}


\bibitem[\protect\citeauthoryear{Guo, Kandel, Hellerstein, and Heer}{Guo et~al\mbox{.}}{2011}]%
        {guoProactiveWranglingMixedinitiative2011}
\bibfield{author}{\bibinfo{person}{Philip~J. Guo}, \bibinfo{person}{Sean Kandel}, \bibinfo{person}{Joseph~M. Hellerstein}, {and} \bibinfo{person}{Jeffrey Heer}.} \bibinfo{year}{2011}\natexlab{}.
\newblock \showarticletitle{Proactive wrangling: mixed-initiative end-user programming of data transformation scripts}. In \bibinfo{booktitle}{\emph{Proceedings of the 24th annual {ACM} symposium on {User} interface software and technology - {UIST} '11}}. \bibinfo{publisher}{ACM Press}, \bibinfo{address}{Santa Barbara, California, USA}, \bibinfo{pages}{65}.
\newblock
\showISBNx{978-1-4503-0716-1}
\urldef\tempurl%
\url{https://doi.org/10.1145/2047196.2047205}
\showDOI{\tempurl}
\newblock
\shownote{00000.}


\bibitem[\protect\citeauthoryear{Heer}{Heer}{2019}]%
        {heerAgencyAutomationDesigning2019}
\bibfield{author}{\bibinfo{person}{Jeffrey Heer}.} \bibinfo{year}{2019}\natexlab{}.
\newblock \showarticletitle{Agency plus automation: {Designing} artificial intelligence into interactive systems}.
\newblock \bibinfo{journal}{\emph{Proceedings of the National Academy of Sciences}} \bibinfo{volume}{116}, \bibinfo{number}{6} (\bibinfo{date}{Feb.} \bibinfo{year}{2019}), \bibinfo{pages}{1844--1850}.
\newblock
\showISSN{0027-8424, 1091-6490}
\urldef\tempurl%
\url{https://doi.org/10.1073/pnas.1807184115}
\showDOI{\tempurl}


\bibitem[\protect\citeauthoryear{Horvitz}{Horvitz}{1999}]%
        {horvitzPrinciplesMixedinitiativeUser1999}
\bibfield{author}{\bibinfo{person}{Eric Horvitz}.} \bibinfo{year}{1999}\natexlab{}.
\newblock \showarticletitle{Principles of {Mixed}-initiative {User} {Interfaces}}. In \bibinfo{booktitle}{\emph{Proceedings of the {SIGCHI} {Conference} on {Human} {Factors} in {Computing} {Systems}}} \emph{(\bibinfo{series}{{CHI} '99})}. \bibinfo{publisher}{ACM}, \bibinfo{address}{New York, NY, USA}, \bibinfo{pages}{159--166}.
\newblock
\showISBNx{978-0-201-48559-2}
\urldef\tempurl%
\url{https://doi.org/10.1145/302979.303030}
\showDOI{\tempurl}


\bibitem[\protect\citeauthoryear{Huang, Quan, Chan, and MacNeil}{Huang et~al\mbox{.}}{2023}]%
        {huangCausalMapperChallengingDesigners2023}
\bibfield{author}{\bibinfo{person}{Ziheng Huang}, \bibinfo{person}{Kexin Quan}, \bibinfo{person}{Joel Chan}, {and} \bibinfo{person}{Stephen MacNeil}.} \bibinfo{year}{2023}\natexlab{}.
\newblock \showarticletitle{{CausalMapper}: {Challenging} designers to think in systems with {Causal} {Maps} and {Large} {Language} {Model}}. In \bibinfo{booktitle}{\emph{Proceedings of the 15th {Conference} on {Creativity} and {Cognition}}} \emph{(\bibinfo{series}{C\&amp;{C} '23})}. \bibinfo{publisher}{Association for Computing Machinery}, \bibinfo{address}{New York, NY, USA}, \bibinfo{pages}{325--329}.
\newblock
\showISBNx{9798400701801}
\urldef\tempurl%
\url{https://doi.org/10.1145/3591196.3596818}
\showDOI{\tempurl}


\bibitem[\protect\citeauthoryear{Jansson and Smith}{Jansson and Smith}{1991}]%
        {janssonDesignFixation1991}
\bibfield{author}{\bibinfo{person}{David~G. Jansson} {and} \bibinfo{person}{Steven~M. Smith}.} \bibinfo{year}{1991}\natexlab{}.
\newblock \showarticletitle{Design fixation}.
\newblock \bibinfo{journal}{\emph{Design Studies}} \bibinfo{volume}{12}, \bibinfo{number}{1} (\bibinfo{year}{1991}), \bibinfo{pages}{3--11}.
\newblock
\urldef\tempurl%
\url{https://doi.org/10.1016/0142-694X(91)90003-F}
\showDOI{\tempurl}


\bibitem[\protect\citeauthoryear{Jiang, Rayan, Dow, and Xia}{Jiang et~al\mbox{.}}{2023}]%
        {jiangGraphologueExploringLarge2023}
\bibfield{author}{\bibinfo{person}{Peiling Jiang}, \bibinfo{person}{Jude Rayan}, \bibinfo{person}{Steven~P. Dow}, {and} \bibinfo{person}{Haijun Xia}.} \bibinfo{year}{2023}\natexlab{}.
\newblock \bibinfo{title}{Graphologue: {Exploring} {Large} {Language} {Model} {Responses} with {Interactive} {Diagrams}}.
\newblock
\newblock
\urldef\tempurl%
\url{https://doi.org/10.1145/3586183.3606737}
\showDOI{\tempurl}
\newblock
\shownote{arXiv:2305.11473 [cs].}


\bibitem[\protect\citeauthoryear{Jun}{Jun}{2024}]%
        {junRTisaneExternalizingConceptual2024}
\bibfield{author}{\bibinfo{person}{Eunice Jun}.} \bibinfo{year}{2024}\natexlab{}.
\newblock \showarticletitle{{rTisane}: {Externalizing} {Conceptual} {Models} for {Data} {Analysis} {Prompts} {Reconsideration} of {Domain} {Assumptions} and {Facilitates} {Statistical} {Modeling}}.
\newblock  (\bibinfo{year}{2024}).
\newblock


\bibitem[\protect\citeauthoryear{Jun, Birchfield, De~Moura, Heer, and Just}{Jun et~al\mbox{.}}{2022}]%
        {junHypothesisFormalizationEmpirical2022}
\bibfield{author}{\bibinfo{person}{Eunice Jun}, \bibinfo{person}{Melissa Birchfield}, \bibinfo{person}{Nicole De~Moura}, \bibinfo{person}{Jeffrey Heer}, {and} \bibinfo{person}{René Just}.} \bibinfo{year}{2022}\natexlab{}.
\newblock \showarticletitle{Hypothesis {Formalization}: {Empirical} {Findings}, {Software} {Limitations}, and {Design} {Implications}}.
\newblock \bibinfo{journal}{\emph{ACM Transactions on Computer-Human Interaction}} \bibinfo{volume}{29}, \bibinfo{number}{1} (\bibinfo{date}{Jan.} \bibinfo{year}{2022}), \bibinfo{pages}{6:1--6:28}.
\newblock
\showISSN{1073-0516}
\urldef\tempurl%
\url{https://doi.org/10.1145/3476980}
\showDOI{\tempurl}


\bibitem[\protect\citeauthoryear{Kahneman}{Kahneman}{2011}]%
        {kahneman2011thinking}
\bibfield{author}{\bibinfo{person}{Daniel Kahneman}.} \bibinfo{year}{2011}\natexlab{}.
\newblock \showarticletitle{Thinking, fast and slow}.
\newblock \bibinfo{journal}{\emph{Farrar, Straus and Giroux}} (\bibinfo{year}{2011}).
\newblock


\bibitem[\protect\citeauthoryear{Kandel, Paepcke, Hellerstein, and Heer}{Kandel et~al\mbox{.}}{2011}]%
        {kandelWranglerInteractiveVisual2011}
\bibfield{author}{\bibinfo{person}{Sean Kandel}, \bibinfo{person}{Andreas Paepcke}, \bibinfo{person}{Joseph Hellerstein}, {and} \bibinfo{person}{Jeffrey Heer}.} \bibinfo{year}{2011}\natexlab{}.
\newblock \showarticletitle{Wrangler: interactive visual specification of data transformation scripts}. In \bibinfo{booktitle}{\emph{{ACM} human factors in computing systems ({CHI})}}.
\newblock
\urldef\tempurl%
\url{https://doi.org/10.1145/1978942.1979444}
\showDOI{\tempurl}


\bibitem[\protect\citeauthoryear{Kazemitabaar, Williams, Drosos, Grossman, Henley, Negreanu, and Sarkar}{Kazemitabaar et~al\mbox{.}}{2024}]%
        {kazemitabaarImprovingSteeringVerification2024}
\bibfield{author}{\bibinfo{person}{Majeed Kazemitabaar}, \bibinfo{person}{Jack Williams}, \bibinfo{person}{Ian Drosos}, \bibinfo{person}{Tovi Grossman}, \bibinfo{person}{Austin~Zachary Henley}, \bibinfo{person}{Carina Negreanu}, {and} \bibinfo{person}{Advait Sarkar}.} \bibinfo{year}{2024}\natexlab{}.
\newblock \showarticletitle{Improving {Steering} and {Verification} in {AI}-{Assisted} {Data} {Analysis} with {Interactive} {Task} {Decomposition}}. In \bibinfo{booktitle}{\emph{Proceedings of the 37th {Annual} {ACM} {Symposium} on {User} {Interface} {Software} and {Technology}}} \emph{(\bibinfo{series}{{UIST} '24})}. \bibinfo{publisher}{Association for Computing Machinery}, \bibinfo{address}{New York, NY, USA}, \bibinfo{pages}{1--19}.
\newblock
\showISBNx{9798400706288}
\urldef\tempurl%
\url{https://doi.org/10.1145/3654777.3676345}
\showDOI{\tempurl}


\bibitem[\protect\citeauthoryear{Lee, Parr, Plaisant, Bederson, Veksler, Gray, and Kotfila}{Lee et~al\mbox{.}}{2006}]%
        {leeTreePlusInteractiveExploration2006}
\bibfield{author}{\bibinfo{person}{Bongshin Lee}, \bibinfo{person}{C.S. Parr}, \bibinfo{person}{C. Plaisant}, \bibinfo{person}{B.B. Bederson}, \bibinfo{person}{V.D. Veksler}, \bibinfo{person}{W.D. Gray}, {and} \bibinfo{person}{C. Kotfila}.} \bibinfo{year}{2006}\natexlab{}.
\newblock \showarticletitle{{TreePlus}: {Interactive} {Exploration} of {Networks} with {Enhanced} {Tree} {Layouts}}.
\newblock \bibinfo{journal}{\emph{IEEE Transactions on Visualization and Computer Graphics}} \bibinfo{volume}{12}, \bibinfo{number}{6} (\bibinfo{date}{Nov.} \bibinfo{year}{2006}), \bibinfo{pages}{1414--1426}.
\newblock
\showISSN{1941-0506}
\urldef\tempurl%
\url{https://doi.org/10.1109/TVCG.2006.106}
\showDOI{\tempurl}


\bibitem[\protect\citeauthoryear{Lee, Gero, Chung, Shum, Raheja, Shen, Venugopalan, Wambsganss, Zhou, Alghamdi, August, Bhat, Choksi, Dutta, Guo, Hoque, Kim, Knight, Neshaei, Sergeyuk, Shibani, Shrivastava, Shroff, Stark, Sterman, Wang, Bosselut, Buschek, Chang, Chen, Kreminski, Park, Pea, Rho, Shen, and Siangliulue}{Lee et~al\mbox{.}}{2024}]%
        {leeDesignSpaceIntelligent2024}
\bibfield{author}{\bibinfo{person}{Mina Lee}, \bibinfo{person}{Katy~Ilonka Gero}, \bibinfo{person}{John Joon~Young Chung}, \bibinfo{person}{Simon~Buckingham Shum}, \bibinfo{person}{Vipul Raheja}, \bibinfo{person}{Hua Shen}, \bibinfo{person}{Subhashini Venugopalan}, \bibinfo{person}{Thiemo Wambsganss}, \bibinfo{person}{David Zhou}, \bibinfo{person}{Emad~A. Alghamdi}, \bibinfo{person}{Tal August}, \bibinfo{person}{Avinash Bhat}, \bibinfo{person}{Madiha~Zahrah Choksi}, \bibinfo{person}{Senjuti Dutta}, \bibinfo{person}{Jin L.~C. Guo}, \bibinfo{person}{Md~Naimul Hoque}, \bibinfo{person}{Yewon Kim}, \bibinfo{person}{Simon Knight}, \bibinfo{person}{Seyed~Parsa Neshaei}, \bibinfo{person}{Agnia Sergeyuk}, \bibinfo{person}{Antonette Shibani}, \bibinfo{person}{Disha Shrivastava}, \bibinfo{person}{Lila Shroff}, \bibinfo{person}{Jessi Stark}, \bibinfo{person}{Sarah Sterman}, \bibinfo{person}{Sitong Wang}, \bibinfo{person}{Antoine Bosselut}, \bibinfo{person}{Daniel Buschek}, \bibinfo{person}{Joseph~Chee Chang},
  \bibinfo{person}{Sherol Chen}, \bibinfo{person}{Max Kreminski}, \bibinfo{person}{Joonsuk Park}, \bibinfo{person}{Roy Pea}, \bibinfo{person}{Eugenia~H. Rho}, \bibinfo{person}{Shannon~Zejiang Shen}, {and} \bibinfo{person}{Pao Siangliulue}.} \bibinfo{year}{2024}\natexlab{}.
\newblock \showarticletitle{A {Design} {Space} for {Intelligent} and {Interactive} {Writing} {Assistants}}. In \bibinfo{booktitle}{\emph{Proceedings of the {CHI} {Conference} on {Human} {Factors} in {Computing} {Systems}}}. \bibinfo{pages}{1--35}.
\newblock
\urldef\tempurl%
\url{https://doi.org/10.1145/3613904.3642697}
\showDOI{\tempurl}
\newblock
\shownote{arXiv:2403.14117 [cs].}


\bibitem[\protect\citeauthoryear{Lewis, Perez, Piktus, Petroni, Karpukhin, Goyal, Küttler, Lewis, Yih, Rocktäschel, Riedel, and Kiela}{Lewis et~al\mbox{.}}{2020}]%
        {lewisRetrievalAugmentedGenerationKnowledgeIntensive2020}
\bibfield{author}{\bibinfo{person}{Patrick Lewis}, \bibinfo{person}{Ethan Perez}, \bibinfo{person}{Aleksandra Piktus}, \bibinfo{person}{Fabio Petroni}, \bibinfo{person}{Vladimir Karpukhin}, \bibinfo{person}{Naman Goyal}, \bibinfo{person}{Heinrich Küttler}, \bibinfo{person}{Mike Lewis}, \bibinfo{person}{Wen-tau Yih}, \bibinfo{person}{Tim Rocktäschel}, \bibinfo{person}{Sebastian Riedel}, {and} \bibinfo{person}{Douwe Kiela}.} \bibinfo{year}{2020}\natexlab{}.
\newblock \showarticletitle{Retrieval-{Augmented} {Generation} for {Knowledge}-{Intensive} {NLP} {Tasks}}. In \bibinfo{booktitle}{\emph{Advances in {Neural} {Information} {Processing} {Systems}}}, Vol.~\bibinfo{volume}{33}. \bibinfo{publisher}{Curran Associates, Inc.}, \bibinfo{pages}{9459--9474}.
\newblock
\urldef\tempurl%
\url{https://proceedings.neurips.cc/paper/2020/hash/6b493230205f780e1bc26945df7481e5-Abstract.html}
\showURL{%
\tempurl}


\bibitem[\protect\citeauthoryear{Manesh, Jr, and Lee}{Manesh et~al\mbox{.}}{2024}]%
        {manesh2024sharp}
\bibfield{author}{\bibinfo{person}{Daniel Manesh}, \bibinfo{person}{Douglas~Bowman Jr}, {and} \bibinfo{person}{Sang~Won Lee}.} \bibinfo{year}{2024}\natexlab{}.
\newblock \showarticletitle{{SHARP}: {Exploring} {Version} {Control} {Systems} in {Live} {Coding} {Music}}.
\newblock  (\bibinfo{year}{2024}).
\newblock


\bibitem[\protect\citeauthoryear{Pirolli and Card}{Pirolli and Card}{1999}]%
        {pirolliInformationForaging1999}
\bibfield{author}{\bibinfo{person}{P. Pirolli} {and} \bibinfo{person}{S. Card}.} \bibinfo{year}{1999}\natexlab{}.
\newblock \showarticletitle{Information foraging}.
\newblock \bibinfo{journal}{\emph{Psychological Review}} \bibinfo{volume}{106}, \bibinfo{number}{4} (\bibinfo{year}{1999}), \bibinfo{pages}{643--675}.
\newblock
\showISSN{0033-295X}
\urldef\tempurl%
\url{https://doi.org/10.1037/0033-295x.106.4.643}
\showDOI{\tempurl}


\bibitem[\protect\citeauthoryear{Plaisant, Grosjean, and Bederson}{Plaisant et~al\mbox{.}}{2002}]%
        {plaisantSpaceTreeSupportingExploration2002}
\bibfield{author}{\bibinfo{person}{C. Plaisant}, \bibinfo{person}{J. Grosjean}, {and} \bibinfo{person}{B.B. Bederson}.} \bibinfo{year}{2002}\natexlab{}.
\newblock \showarticletitle{{SpaceTree}: supporting exploration in large node link tree, design evolution and empirical evaluation}. In \bibinfo{booktitle}{\emph{{IEEE} {Symposium} on {Information} {Visualization}, 2002. {INFOVIS} 2002.}} \bibinfo{publisher}{IEEE Comput. Soc}, \bibinfo{address}{Boston, MA, USA}, \bibinfo{pages}{57--64}.
\newblock
\showISBNx{978-0-7695-1751-3}
\urldef\tempurl%
\url{https://doi.org/10.1109/INFVIS.2002.1173148}
\showDOI{\tempurl}


\bibitem[\protect\citeauthoryear{Quadri}{Quadri}{2024}]%
        {quadriYouSeeWhat2024}
\bibfield{author}{\bibinfo{person}{Ghulam~Jilani Quadri}.} \bibinfo{year}{2024}\natexlab{}.
\newblock \showarticletitle{Do {You} {See} {What} {I} {See}? {A} {Qualitative} {Study} {Eliciting} {High}-{Level} {Visualization} {Comprehension}}.
\newblock  (\bibinfo{year}{2024}).
\newblock


\bibitem[\protect\citeauthoryear{Rittberg, Tanswell, and Van~Bendegem}{Rittberg et~al\mbox{.}}{2018}]%
        {rittbergEpistemicInjusticeMathematics2018}
\bibfield{author}{\bibinfo{person}{Colin~Jakob Rittberg}, \bibinfo{person}{Fenner~Stanley Tanswell}, {and} \bibinfo{person}{Jean~Paul Van~Bendegem}.} \bibinfo{year}{2018}\natexlab{}.
\newblock \showarticletitle{Epistemic injustice in mathematics}.
\newblock \bibinfo{journal}{\emph{Synthese}} (\bibinfo{date}{Oct.} \bibinfo{year}{2018}).
\newblock
\showISSN{1573-0964}
\urldef\tempurl%
\url{https://doi.org/10.1007/s11229-018-01981-1}
\showDOI{\tempurl}


\bibitem[\protect\citeauthoryear{Satyanarayan, Moritz, Wongsuphasawat, and Heer}{Satyanarayan et~al\mbox{.}}{[n.d.]}]%
        {satyanarayanVegaLiteGrammarInteractive}
\bibfield{author}{\bibinfo{person}{Arvind Satyanarayan}, \bibinfo{person}{Dominik Moritz}, \bibinfo{person}{Kanit Wongsuphasawat}, {and} \bibinfo{person}{Jeffrey Heer}.} \bibinfo{year}{[n.d.]}\natexlab{}.
\newblock \showarticletitle{Vega-{Lite}: {A} {Grammar} of {Interactive} {Graphics}}.
\newblock  (\bibinfo{year}{[n.\,d.]}).
\newblock


\bibitem[\protect\citeauthoryear{Si, Yang, and Hashimoto}{Si et~al\mbox{.}}{2024}]%
        {siCanLLMsGenerate2024}
\bibfield{author}{\bibinfo{person}{Chenglei Si}, \bibinfo{person}{Diyi Yang}, {and} \bibinfo{person}{Tatsunori Hashimoto}.} \bibinfo{year}{2024}\natexlab{}.
\newblock \bibinfo{title}{Can {LLMs} {Generate} {Novel} {Research} {Ideas}? {A} {Large}-{Scale} {Human} {Study} with 100+ {NLP} {Researchers}}.
\newblock
\newblock
\urldef\tempurl%
\url{http://arxiv.org/abs/2409.04109}
\showURL{%
\tempurl}
\newblock
\shownote{arXiv:2409.04109 [cs].}


\bibitem[\protect\citeauthoryear{team, Barrault, Duquenne, Elbayad, Kozhevnikov, Alastruey, Andrews, Coria, Couairon, Costa-jussà, Dale, Elsahar, Heffernan, Janeiro, Tran, Ropers, Sánchez, Roman, Mourachko, Saleem, and Schwenk}{team et~al\mbox{.}}{2024}]%
        {teamLargeConceptModels2024}
\bibfield{author}{\bibinfo{person}{L.~C.~M. team}, \bibinfo{person}{Loïc Barrault}, \bibinfo{person}{Paul-Ambroise Duquenne}, \bibinfo{person}{Maha Elbayad}, \bibinfo{person}{Artyom Kozhevnikov}, \bibinfo{person}{Belen Alastruey}, \bibinfo{person}{Pierre Andrews}, \bibinfo{person}{Mariano Coria}, \bibinfo{person}{Guillaume Couairon}, \bibinfo{person}{Marta~R. Costa-jussà}, \bibinfo{person}{David Dale}, \bibinfo{person}{Hady Elsahar}, \bibinfo{person}{Kevin Heffernan}, \bibinfo{person}{João~Maria Janeiro}, \bibinfo{person}{Tuan Tran}, \bibinfo{person}{Christophe Ropers}, \bibinfo{person}{Eduardo Sánchez}, \bibinfo{person}{Robin~San Roman}, \bibinfo{person}{Alexandre Mourachko}, \bibinfo{person}{Safiyyah Saleem}, {and} \bibinfo{person}{Holger Schwenk}.} \bibinfo{year}{2024}\natexlab{}.
\newblock \bibinfo{title}{Large {Concept} {Models}: {Language} {Modeling} in a {Sentence} {Representation} {Space}}.
\newblock
\newblock
\urldef\tempurl%
\url{https://doi.org/10.48550/arXiv.2412.08821}
\showDOI{\tempurl}
\newblock
\shownote{arXiv:2412.08821 [cs].}


\bibitem[\protect\citeauthoryear{Tian, Cui, Deng, Yi, Yang, Zhang, and Wu}{Tian et~al\mbox{.}}{2024}]%
        {tianChartGPTLeveragingLLMs2024}
\bibfield{author}{\bibinfo{person}{Yuan Tian}, \bibinfo{person}{Weiwei Cui}, \bibinfo{person}{Dazhen Deng}, \bibinfo{person}{Xinjing Yi}, \bibinfo{person}{Yurun Yang}, \bibinfo{person}{Haidong Zhang}, {and} \bibinfo{person}{Yingcai Wu}.} \bibinfo{year}{2024}\natexlab{}.
\newblock \showarticletitle{{ChartGPT}: {Leveraging} {LLMs} to {Generate} {Charts} from {Abstract} {Natural} {Language}}.
\newblock \bibinfo{journal}{\emph{IEEE Transactions on Visualization and Computer Graphics}} (\bibinfo{year}{2024}), \bibinfo{pages}{1--15}.
\newblock
\showISSN{1077-2626, 1941-0506, 2160-9306}
\urldef\tempurl%
\url{https://doi.org/10.1109/TVCG.2024.3368621}
\showDOI{\tempurl}
\newblock
\shownote{arXiv:2311.01920 [cs].}


\bibitem[\protect\citeauthoryear{Tong, Mao, Huang, Zhao, and Peng}{Tong et~al\mbox{.}}{2024}]%
        {tong2024automating}
\bibfield{author}{\bibinfo{person}{Song Tong}, \bibinfo{person}{Kai Mao}, \bibinfo{person}{Zhen Huang}, \bibinfo{person}{Yukun Zhao}, {and} \bibinfo{person}{Kaiping Peng}.} \bibinfo{year}{2024}\natexlab{}.
\newblock \showarticletitle{Automating psychological hypothesis generation with {AI}: when large language models meet causal graph}.
\newblock \bibinfo{journal}{\emph{Humanities and Social Sciences Communications}} \bibinfo{volume}{11}, \bibinfo{number}{1} (\bibinfo{date}{July} \bibinfo{year}{2024}), \bibinfo{pages}{896}.
\newblock
\showISSN{2662-9992}
\urldef\tempurl%
\url{https://doi.org/10.1057/s41599-024-03407-5}
\showDOI{\tempurl}


\bibitem[\protect\citeauthoryear{Vaithilingam, Glassman, Inala, and Wang}{Vaithilingam et~al\mbox{.}}{2024}]%
        {vaithilingamDynaVisDynamicallySynthesized2024}
\bibfield{author}{\bibinfo{person}{Priyan Vaithilingam}, \bibinfo{person}{Elena~L. Glassman}, \bibinfo{person}{Jeevana~Priya Inala}, {and} \bibinfo{person}{Chenglong Wang}.} \bibinfo{year}{2024}\natexlab{}.
\newblock \showarticletitle{{DynaVis}: {Dynamically} {Synthesized} {UI} {Widgets} for {Visualization} {Editing}}. In \bibinfo{booktitle}{\emph{Proceedings of the {CHI} {Conference} on {Human} {Factors} in {Computing} {Systems}}} \emph{(\bibinfo{series}{{CHI} '24})}. \bibinfo{publisher}{Association for Computing Machinery}, \bibinfo{address}{New York, NY, USA}, \bibinfo{pages}{1--17}.
\newblock
\showISBNx{9798400703300}
\urldef\tempurl%
\url{https://doi.org/10.1145/3613904.3642639}
\showDOI{\tempurl}


\bibitem[\protect\citeauthoryear{Weidele, Martino, Valente, Rossiello, Strobelt, Franke, Alvero, Misko, Auer, Bagchi, Mihindukulasooriya, Chowdhury, Bramble, Samulowitz, Gliozzo, and Amini}{Weidele et~al\mbox{.}}{2024}]%
        {weideleEmpiricalEvidenceConversational2024}
\bibfield{author}{\bibinfo{person}{Daniel Karl~I. Weidele}, \bibinfo{person}{Mauro Martino}, \bibinfo{person}{Abel~N. Valente}, \bibinfo{person}{Gaetano Rossiello}, \bibinfo{person}{Hendrik Strobelt}, \bibinfo{person}{Loraine Franke}, \bibinfo{person}{Kathryn Alvero}, \bibinfo{person}{Shayenna Misko}, \bibinfo{person}{Robin Auer}, \bibinfo{person}{Sugato Bagchi}, \bibinfo{person}{Nandana Mihindukulasooriya}, \bibinfo{person}{Faisal Chowdhury}, \bibinfo{person}{Gregory Bramble}, \bibinfo{person}{Horst Samulowitz}, \bibinfo{person}{Alfio Gliozzo}, {and} \bibinfo{person}{Lisa Amini}.} \bibinfo{year}{2024}\natexlab{}.
\newblock \showarticletitle{Empirical {Evidence} on {Conversational} {Control} of {GUI} in {Semantic} {Automation}}. In \bibinfo{booktitle}{\emph{Proceedings of the 29th {International} {Conference} on {Intelligent} {User} {Interfaces}}}. \bibinfo{publisher}{ACM}, \bibinfo{address}{Greenville SC USA}, \bibinfo{pages}{869--885}.
\newblock
\showISBNx{9798400705083}
\urldef\tempurl%
\url{https://doi.org/10.1145/3640543.3645172}
\showDOI{\tempurl}


\bibitem[\protect\citeauthoryear{Wongsuphasawat, Moritz, Anand, Mackinlay, Howe, and Heer}{Wongsuphasawat et~al\mbox{.}}{2016}]%
        {wongsuphasawatVoyagerExploratoryAnalysis2016}
\bibfield{author}{\bibinfo{person}{Kanit Wongsuphasawat}, \bibinfo{person}{Dominik Moritz}, \bibinfo{person}{Anushka Anand}, \bibinfo{person}{Jock Mackinlay}, \bibinfo{person}{Bill Howe}, {and} \bibinfo{person}{Jeffrey Heer}.} \bibinfo{year}{2016}\natexlab{}.
\newblock \showarticletitle{Voyager: {Exploratory} {Analysis} via {Faceted} {Browsing} of {Visualization} {Recommendations}}.
\newblock \bibinfo{journal}{\emph{IEEE Transactions on Visualization and Computer Graphics}} \bibinfo{volume}{22}, \bibinfo{number}{1} (\bibinfo{date}{Jan.} \bibinfo{year}{2016}), \bibinfo{pages}{649--658}.
\newblock
\showISSN{1077-2626}
\urldef\tempurl%
\url{https://doi.org/10.1109/TVCG.2015.2467191}
\showDOI{\tempurl}


\bibitem[\protect\citeauthoryear{Wongsuphasawat, Qu, Moritz, Chang, Ouk, Anand, Mackinlay, Howe, and Heer}{Wongsuphasawat et~al\mbox{.}}{2017}]%
        {wongsuphasawatVoyagerAugmentingVisual2017}
\bibfield{author}{\bibinfo{person}{Kanit Wongsuphasawat}, \bibinfo{person}{Zening Qu}, \bibinfo{person}{Dominik Moritz}, \bibinfo{person}{Riley Chang}, \bibinfo{person}{Felix Ouk}, \bibinfo{person}{Anushka Anand}, \bibinfo{person}{Jock Mackinlay}, \bibinfo{person}{Bill Howe}, {and} \bibinfo{person}{Jeffrey Heer}.} \bibinfo{year}{2017}\natexlab{}.
\newblock \showarticletitle{Voyager 2: {Augmenting} {Visual} {Analysis} with {Partial} {View} {Specifications}}. In \bibinfo{booktitle}{\emph{Proceedings of the 2017 {CHI} {Conference} on {Human} {Factors} in {Computing} {Systems}}}. \bibinfo{publisher}{ACM}, \bibinfo{address}{Denver Colorado USA}, \bibinfo{pages}{2648--2659}.
\newblock
\showISBNx{978-1-4503-4655-9}
\urldef\tempurl%
\url{https://doi.org/10.1145/3025453.3025768}
\showDOI{\tempurl}


\bibitem[\protect\citeauthoryear{Xia}{Xia}{2024}]%
        {xiaRedesigningInformationSpace2024}
\bibfield{author}{\bibinfo{person}{Haijun Xia}.} \bibinfo{year}{2024}\natexlab{}.
\newblock \showarticletitle{Redesigning the {Information} {Space} to {Unleash} the {Power} of {AI}}.
\newblock  (\bibinfo{year}{2024}).
\newblock
\urldef\tempurl%
\url{https://creativity.ucsd.edu/foundation}
\showURL{%
\tempurl}


\bibitem[\protect\citeauthoryear{Xie, Zheng, Xia, Qu, and Zhu-Tian}{Xie et~al\mbox{.}}{2024}]%
        {xieWaitGPTMonitoringSteering2024}
\bibfield{author}{\bibinfo{person}{Liwenhan Xie}, \bibinfo{person}{Chengbo Zheng}, \bibinfo{person}{Haijun Xia}, \bibinfo{person}{Huamin Qu}, {and} \bibinfo{person}{Chen Zhu-Tian}.} \bibinfo{year}{2024}\natexlab{}.
\newblock \bibinfo{title}{{WaitGPT}: {Monitoring} and {Steering} {Conversational} {LLM} {Agent} in {Data} {Analysis} with {On}-the-{Fly} {Code} {Visualization}}.
\newblock
\newblock
\urldef\tempurl%
\url{https://doi.org/10.1145/3654777.3676374}
\showDOI{\tempurl}
\newblock
\shownote{arXiv:2408.01703 [cs].}


\bibitem[\protect\citeauthoryear{Zhang, Hendra, Chi, and Ding}{Zhang et~al\mbox{.}}{2024}]%
        {zhangFrontendDiffusionExploring2024}
\bibfield{author}{\bibinfo{person}{Qinshi Zhang}, \bibinfo{person}{Latisha~Besariani Hendra}, \bibinfo{person}{Mohan Chi}, {and} \bibinfo{person}{Zijian Ding}.} \bibinfo{year}{2024}\natexlab{}.
\newblock \bibinfo{title}{Frontend {Diffusion}: {Exploring} {Intent}-{Based} {User} {Interfaces} through {Abstract}-to-{Detailed} {Task} {Transitions}}.
\newblock
\newblock
\urldef\tempurl%
\url{http://arxiv.org/abs/2408.00778}
\showURL{%
\tempurl}
\newblock
\shownote{arXiv:2408.00778 [cs].}


\bibitem[\protect\citeauthoryear{Zheng, Wang, Wang, and Ma}{Zheng et~al\mbox{.}}{2022}]%
        {zheng2022telling}
\bibfield{author}{\bibinfo{person}{Chengbo Zheng}, \bibinfo{person}{Dakuo Wang}, \bibinfo{person}{April~Yi Wang}, {and} \bibinfo{person}{Xiaojuan Ma}.} \bibinfo{year}{2022}\natexlab{}.
\newblock \showarticletitle{Telling {Stories} from {Computational} {Notebooks}: {AI}-{Assisted} {Presentation} {Slides} {Creation} for {Presenting} {Data} {Science} {Work}}. In \bibinfo{booktitle}{\emph{{CHI} {Conference} on {Human} {Factors} in {Computing} {Systems}}}. \bibinfo{publisher}{ACM}, \bibinfo{address}{New Orleans LA USA}, \bibinfo{pages}{1--20}.
\newblock
\showISBNx{978-1-4503-9157-3}
\urldef\tempurl%
\url{https://doi.org/10.1145/3491102.3517615}
\showDOI{\tempurl}


\bibitem[\protect\citeauthoryear{Zhou, Liu, Srivastava, Mei, and Tan}{Zhou et~al\mbox{.}}{2024}]%
        {zhouHypothesisGenerationLarge2024}
\bibfield{author}{\bibinfo{person}{Yangqiaoyu Zhou}, \bibinfo{person}{Haokun Liu}, \bibinfo{person}{Tejes Srivastava}, \bibinfo{person}{Hongyuan Mei}, {and} \bibinfo{person}{Chenhao Tan}.} \bibinfo{year}{2024}\natexlab{}.
\newblock \bibinfo{title}{Hypothesis {Generation} with {Large} {Language} {Models}}.
\newblock
\newblock
\urldef\tempurl%
\url{http://arxiv.org/abs/2404.04326}
\showURL{%
\tempurl}
\newblock
\shownote{arXiv:2404.04326 [cs].}


\end{thebibliography}


\newpage

\appendix

\onecolumn

\section{Appendix A: Prompts for Intent-Hypothesis Interpreter}
\label{appendix:A}

\subsection{Prompt for Initial Hypothesis Generation}

The prompt for initial hypothesis generation was adapted from the LIDA framework \cite{dibiaLIDAToolAutomatic2023}. Our approach extends LIDA's capabilities by specifically focusing on hypothesis generation. Below is structure of the API call and the content for each component:

\begin{verbatim}
messages = [
    {"role": "system", "content": SYSTEM_INSTRUCTIONS},
    {"role": "assistant",
     "content":
     f"{USER_PROMPT}\n\n {FORMAT_INSTRUCTIONS} \n\n. The generated {n} goals are: \n "}]

SYSTEM_INSTRUCTIONS:
You are an experienced data analyst who can generate a given number of insightful Hypothesis 
about data, when given a summary of the data, and a specified persona. The VISUALIZATIONS YOU 
RECOMMEND MUST FOLLOW VISUALIZATION BEST PRACTICES (e.g., must use bar charts instead of pie 
charts for comparing quantities) AND BE MEANINGFUL (e.g., plot longitude and latitude on maps 
where appropriate). They must also be relevant to the specified persona. Each goal must include 
a hypothesis with a title (in [], and the title don't need include the target variable, just 
include the new variable in the hypothesis), a visualization (THE VISUALIZATION MUST REFERENCE 
THE EXACT COLUMN FIELDS FROM THE SUMMARY), and a rationale (JUSTIFICATION FOR WHICH dataset 
FIELDS ARE USED and what we will learn from the visualization).

USER_PROMPT:
The number of Hypothesis to generate is 5. The goals should be based on the data summary below, 
{data summary}. The generated Hypothesis SHOULD BE FOCUSED ON THE INTERESTS AND PERSPECTIVE of 
a {persona.persona} persona, who is interested in complex, insightful goals about the data.

FORMAT_INSTRUCTIONS:
THE OUTPUT MUST BE A CODE SNIPPET OF A VALID LIST OF JSON OBJECTS. IT MUST USE THE FOLLOWING 
FORMAT:
[
    { "index": 0,  "hypothesis": "[new variable X]: Y is highly correlated to X.", 
      "visualization": "scatterplot of X and Y", 
      "rationale": "This tells about "} ..
]
THE OUTPUT SHOULD ONLY USE THE JSON FORMAT ABOVE.
\end{verbatim}

\subsection{New Hypothesis Branch Generation (Hypothesis Iteration)}

\begin{verbatim}
Based on the hypothesis {selected hypothesis} [and the user input {user input}](optional, if 
there is user input to guide new hypothesis generation), generate 3 new and more insightful 
hypotheses based on the given hypothesis. Format the output as a JSON array with the following 
structure for each hypothesis:
{
  "title": "short new variable X (no more than two words)",
  "hypothesis": "There is a...",
  "relatedWork": "Previous studies have shown that...",
  "visualization": "Description of visualization idea",
  "rationale": "Rationale for the visualization"
}
\end{verbatim}

\clearpage

\section{Appendix B: Table of User Engagement with Visual Hypotheses}
\label{appendix:B}

\begin{table}[h]
    \centering
    \begin{tabular}{|c|p{5cm}|p{4.5cm}|p{1.5cm}|}
        \hline
        ID & Initial expansion of visual hypotheses & Re-expansion of visual hypotheses & Total \\
        \hline
        Mean (Variance) & 6.32 (4.90)  & 1.91 (3.85) & 8.23 (7.41) \\
        \hline
        P1 & 18 & 5 & 23 \\
        \hline
        P2 & 5 & 4 & 9 \\
        \hline
        P3 & 13 & 3 & 16 \\
        \hline
        P4 & 9 & 1 & 10 \\
        \hline
        P5 & 9 & 2 & 11 \\
        \hline
        P6 & 5 & 1 & 6 \\
        \hline
        P7 & 7 & 2 & 9 \\
        \hline
        P8 & 8 & 2 & 10 \\
        \hline
        P9 & 2 & 0 & 2 \\
        \hline
        P10 & 11 & 0 & 11 \\
        \hline
        P11 & 14 & 1 & 15 \\
        \hline
        P12 & 1 & 1 & 2 \\
        \hline
        P13 & 0 & 0 & 0 \\
\hline
P14 & 2 & 0 & 2 \\
\hline
P15 & 2 & 0 & 2 \\
\hline
P16 & 3 & 0 & 3 \\
\hline
P17 & 2 & 0 & 2 \\
\hline
P18 & 6 & 1 & 7 \\
\hline
P19 & 11 & 18 & 29 \\
\hline
P20 & 4 & 1 & 5 \\
\hline
P21 & 0 & 0 & 0 \\
\hline
P22 & 7 & 0 & 7 \\
\hline
    \end{tabular}
    \caption{Participant breakdown of user engagement with visual hypotheses.}
    \label{tab:engagement}
\end{table}

\clearpage

\section{Appendix C: Backtrack Behaviors}
\label{appendix:C}

\begin{figure}[h]
\includegraphics[width=0.8\textwidth]{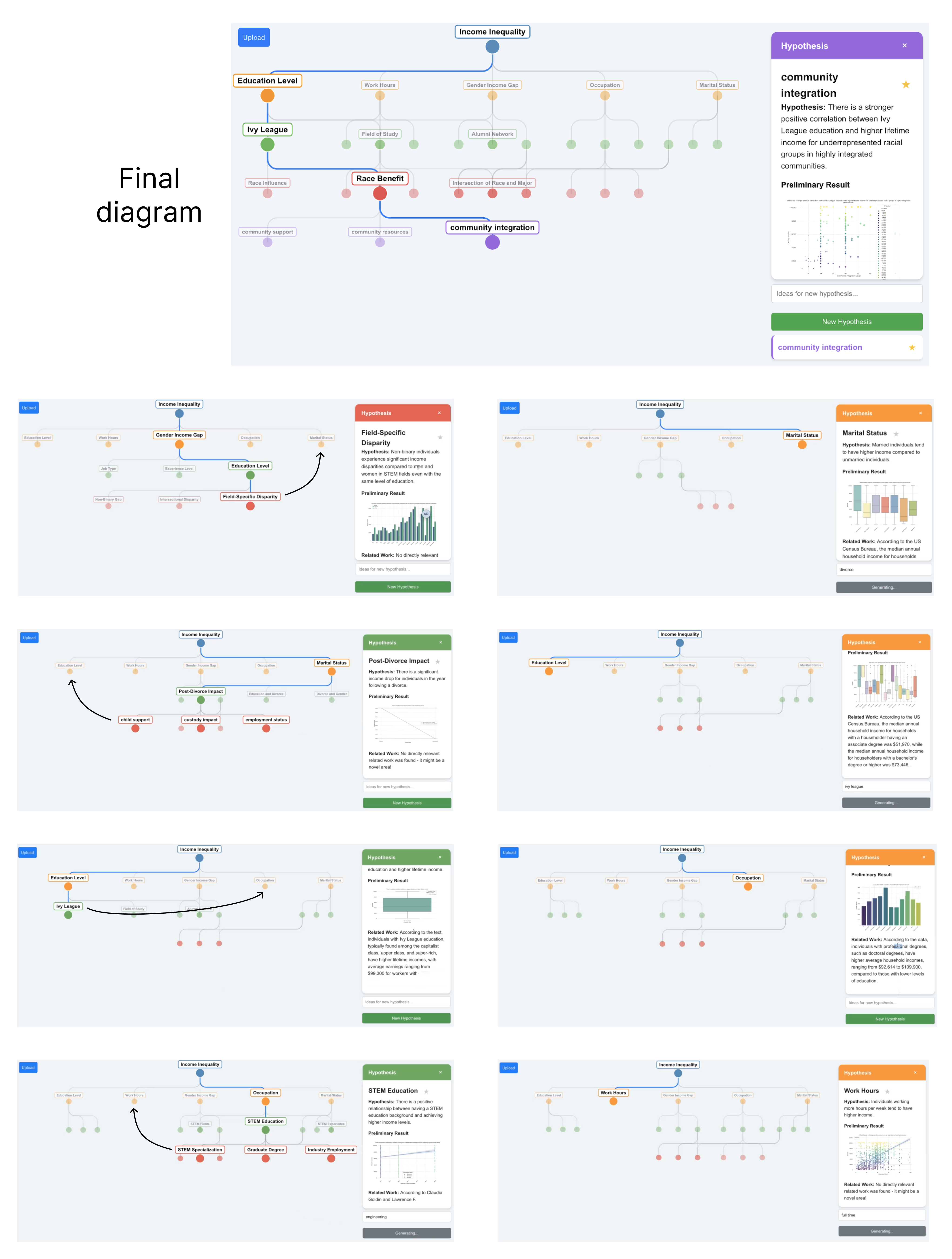}
  \caption{Backtrack behaviors of P3 (17 backtracks in total). Part of it was covered in Figure \ref{fig:teaser}.}
  \Description{}
  \label{fig:P3}
\end{figure}

\begin{table}[h]
    \centering
    \begin{tabular}{|c|p{4.5cm}|p{3.5cm}|p{2.2cm}|p{1.5cm}|}
        \hline
        ID & High Level Backtrack and Generate & High Level Backtrack Only & Other Backtrack & Total \\
        \hline
        Mean (Variance) & 0.91 (1.52) & 1.64 (2.15) & 1.41 (5.49) & 3.95 (12.90) \\
        \hline
        P1 & 0 & 2 & 0 & 2 \\
        \hline
        P2 & 0 & 3 & 3 & 6 \\
        \hline
        P3 & 5 & 1 & 11 & 17 \\
        \hline
        P4 & 1 & 3 & 2 & 6 \\
        \hline
        P5 & 1 & 3 & 1 & 5 \\
        \hline
        P6 & 0 & 3 & 0 & 3 \\
        \hline
        P7 & 1 & 3 & 1 & 5 \\
        \hline
        P8 & 2 & 1 & 1 & 4 \\
        \hline
        P9 & 3 & 0 & 2 & 5 \\
        \hline
        P10 & 0 & 1 & 1 & 2 \\
        \hline
        P11 & 1 & 0 & 0 & 1 \\
        \hline
        P12 & 1 & 1 & 0 & 2 \\
        \hline
        P13 & 0 & 1 & 0 & 1 \\
\hline
P14 & 0 & 0 & 0 & 0 \\
\hline
P15 & 2 & 0 & 2 & 4 \\
\hline
P16 & 1 & 1 & 1 & 3 \\
\hline
P17 & 0 & 4 & 2 & 6 \\
\hline
P18 & 0 & 5 & 2 & 7 \\
\hline
P19 & 0 & 2 & 0 & 2 \\
\hline
P20 & 0 & 0 & 0 & 0 \\
\hline
P21 & 1 & 2 & 2 & 5 \\
\hline
P22 & 1 & 0 & 0 & 1 \\
\hline
    \end{tabular}
    \caption{Participant breakdown of three types of backtrack behaviors: high level backtrack and generate (revisit a hypothesis node on a higher level and not on the trajectory from the current node to the root node, and generate new branch from the node), high level backtrack only (only revisit a hypothesis node on a high level and not on the current trajectory) and other backtrack (revisit a hypothesis node but not on a higher level).}
    \label{tab:backtrack}
\end{table}

\end{document}